\documentclass{emulateapj}
\usepackage{placeins}
\usepackage{graphicx}
\shorttitle{Particles Trapped at Gap Edges and Vortices} \shortauthors{Zhu et al.}

\newcommand\msunyr{\rm M_{\odot}\,yr^{-1}}
\newcommand\be{\begin{equation}}
\newcommand\en{\end{equation}}

\newcommand\etal{{\rm et al}.\ }

\begin{document}

\title{ Particle Concentration At Planet Induced Gap Edges and Vortices : I. Inviscid 3-D Hydro Disks}

\author{Zhaohuan Zhu\altaffilmark{1},  James M. Stone\altaffilmark{1}, Roman R. Rafikov\altaffilmark{1}, Xuening Bai\altaffilmark{2} }

\altaffiltext{1}{Department of Astrophysical Sciences, 4 Ivy Lane, Peyton Hall,
Princeton University, Princeton, NJ 08544}
\altaffiltext{2}{Center for Astrophysics,
60 Garden St., Cambridge, MA 02138}

\email{zhzhu@astro.princeton.edu }

\begin{abstract}
We perform a systematic study of the dynamics of dust particles in protoplanetary disks with embedded planets using global 2-D and 3-D inviscid hydrodynamic simulations. Lagrangian particles have been
implemented into magnetohydrodynamic code Athena with cylindrical coordinates.
We find two distinct outcomes depending on the mass of the embedded planet. In the presence of a low mass planet ($8 M_{\oplus}$), two narrow gaps start to open in the gas on each side of the planet where the density waves shock. These shallow gaps can dramatically affect particle drift speed and cause significant, roughly axisymmetric dust depletion. On the other hand, a more massive planet ($>0.1 M_{J}$) carves out a deeper gap with sharp edges, which are unstable to the vortex formation. Particles with a wide range of sizes ($0.02<\Omega t_{s}<20$)
are trapped and settle to the midplane in the vortex, with the strongest concentration for particles with $\Omega t_{s}\sim 1$.
The dust concentration is highly elongated in the $\phi$ direction, and can be
as wide as 4 disk scale heights in the radial direction. 
Dust surface density inside the vortex can be increased by more than 
a factor of 10$^2$  in a very non-axisymmetric fashion. For very big particles ($\Omega t_{s}\gg 1$) we find strong eccentricity excitation, in particular around the planet and in the vicinity of the mean motion resonances, facilitating gap opening there. Our results imply that in  weakly turbulent protoplanetary disk regions (e.g. the ``dead zone'') dust particles with a very wide range of sizes can be trapped at gap edges and inside vortices induced by planets with $M_{p}<M_{J}$, potentially accelerating planetesimal and planet formation there, and giving rise to distinctive features that can be probed by ALMA and EVLA.
\end{abstract}

\keywords{accretion disks, stars: formation, stars: pre-main
sequence}

\section{Introduction}
The
gravitational interaction of a planet with a protoplanetary disk
 can significantly change the evolution of both the planet and disk.
Planets can migrate in disks through the excitation of density waves 
(Goldreich \& Tremaine 1979, Tanaka \etal 2002).
On the other hand, the planet can open a gap in
the disk if angular momentum carried by density waves can be efficiently
 transferred to the disk
 (Lin \& Papaloizou 1986). 

Recently,
there have been two major developments in our understanding of 
the gap opening process. The first 
is the realization that low mass planets can also open gaps through
nonlinear density wave steepening
(Goodman \& Rafikov 2001, Rafikov 2002a). 
Traditionally, it had been thought that the density wave needs to be highly
nonlinear at the location where it is excited in order to open a gap around the planet (Lin \& Papaloizou 1993). 
This translates to a criterion that, to open gaps, the planet mass needs to be above 
the thermal mass defined as
\begin{equation}
M_{th}=\frac{c_{s}^{3}}{G\Omega_{p}}=M_{*}\left(\frac{H}{R}\right)^{3}\,,\label{eq:thermal}
\end{equation}
where $c_{s}$ and $H$ are the local sound speed and scale height at the planet's position. With
$H/R=0.05$, $M_{th}=0.13 M_{J}$.
However, nonlinear density wave theory (Goodman \& Rafikov 2001) suggests that even if the planet
mass is below the thermal mass and the density
waves are linear at excitation, they gradually but efficiently steepen, ultimately turning into
weak shocks not too far from the planet. Dissipation of the wave angular momentum at these
shocks drives gap opening even around low mass planets (Rafikov 2002b). This picture
has been confirmed recently by high resolution simulations (Li \etal 2009, Muto \etal 2010, 
Dong \etal 2011b, Duffell \& MacFadyen 2012,  Zhu \etal 2013). Gap opening by low mass planets can 
significantly slow down planet migration (Rafikov 2002b, Li \etal 2009), which has far-reaching implications for the
survivability of planets. Furthermore, the fact that gaps can be induced by low mass planets also suggests that gaps 
can be very common in protoplanetary disks, which may explain the recent observations of protoplanetary disks
with gaps and cavities.

The second intriguing development is that, when the disk viscosity is low enough ($\nu<10^{-5}R^{2}\Omega$),
vortices are generated at gap edges (Koller, Li \& Lin 2003, Li \etal 2005, de Val-Borro et 
al. 2007, Lin \& Papaloizou 2010). These vortices are produced by the Rossby 
wave instability (RWI) (Lovelace \etal 1999, Li \etal 2000).  At the planet-induced gap edges, 
spiral shocks introduce 
vortensity minima which lead to RWI and  vortex formation. 
Several vortices form initially, which then quickly merge into one vortex. The interaction between the planet and
the vortex can ultimately affect 
planet migration (Li \etal 2009, Yu \etal 2010, Lin \& Papaloizou 2010).

The interaction between the planet and the gaseous 
disk has been studied systematically in many papers (see
the  review by Kley \& Nelson 2012 for references). However, the 
effect of a planet on dust particles and planetesimals in protoplanetary disks is less well
studied. Although the total mass of dust is only 1/100 of the gas mass,
it represents the key element for building planets and can be detected much more easily since
the opacity of the dust is orders of magnitude larger than that of the gas.
In this paper, we explore how these two recent developments
on gap opening will manifest themselves in dust particles.

Although a low mass planet (e.g. 10 $M_{\oplus}$) takes a long time  to open 
noticeable gaps in the gas, even a very shallow gap in the gas during the gap
opening process will change the drift speed of dust particles significantly
 (Whipple 1972, Weidenschilling 1977). 
Particles tend to accumulate at gap edges and produce a far more
significant particle gap. Inside this gap, the planet will accrete gas
having little dust, and the metallicity of the planet may be lowered.
Such particle concentrations at gap edges can also facilitate planetesimal
formation. Coincidentally, the fastest drifting dust particles, which have
the most prominent dust gaps, are in the submm-cm
size range when the particle is at $R\sim 10-100$ AU. This provides an opportunity
for submm (e.g. ALMA) and cm (e.g. EVLA) observations to probe these gaps.

When a sharp gap is carved out by the planet, either slowly by a low mass planet
or quickly by a massive planet, the gap edges start to develop 
vortices. Vortices are ideal places to trap dust particles
(Barge \& Sommeria 1995, Lyra \etal 2009ab, Johansen \etal 2004; Heng \& Kenyon 2010, Meheut 2012). The three
 outstanding 
issues are (1) can vortices survive in 3-D? (2) particles of what size can accumulate in the vortices? and (3) how 
are particles distributed in the vortices? A 3-D vortex can undergo elliptical instability (Lesur \& Papaloizou 2009).
Even if the vortex survive for a long period of time, it may not concentrate a large amount of dust
if only particles in a narrow size range can accumulate
within the vortex. If the vortex has strong
vertical motions which can stir up dust particles, strong settling towards the
disk midplane is unlikely, inhibiting planetesimal formation within the vortex.
To answer these questions, we need to understand the dynamics and 3-D structure
of the vortices. It requires evolving the particle distribution together
with the gas. 

Evolution of dust particles in protoplanetary disks needs to be calculated by  directly
integrating the orbits of a large number of dust particles
or solving the collisionless-Boltzman equation for the particle distribution function.
Both approaches are quite different from solving the Navier-Stokes equations for the gas. 
In some limiting cases, the Boltzman equation can be reduced to a simpler form.
For example, when the particles are very small, 
their coupling timescale with the gas is far smaller than the orbital timescale. 
In this case, a simple closure can be found, reducing 
the Boltzman equation to the zero pressure fluid equation (Garaud \etal 2004). 

Some initial attempts using this zero-pressure fluid approach to study
particle response to the planet and gap in the gas have already been made
(Paardekooper \& Mellema 2004, 2006, Zhu et al 2012). However, these 2-D two-fluid simulations
were limited to small particles. 

In order to study particles of any size, we need to treat dust particles as individual 
particles and integrate their orbits with N-body integrators.   
SPH codes are ideal for this task by design and they have
been used to study planet's effects on dust particles (Fouchet \etal 2007, 2010, Ayliffe 
\etal 2012). However, for the gas, large viscosities are normally
introduced in SPH codes, so that they cannot be used to study gap opening
by low mass planets and RWI at gap edges.
In order to treat gas dynamics accurately and particle dynamics self-consistently,
we need to incorporate an N-body code into a grid-based fluid code. Lyra \etal (2009a)
have developed such a hybrid scheme to simulate gap opening in global disks using 
2-D Cartesian grids and pointed out that particles are trapped in the vortices. 
However, they also noticed that
RWI is not reproduced well at the gap edge when the gap is shallow.
They suggest that Cartesian
grids need higher resolutions than the cylindrical grids 
to capture the formation of the gap edge vortices, and solving 
 fluid equations and integrating particle orbits under cylindrical coordinates 
 may be more appropriate (also
 pointed out by de Val-Borro \etal 2007). 

In this paper, we develop and implement three particle integrators
using cylindrical coordinates  into
magnetohydrodynamic code Athena 
(Stone \etal 2008). The particle integrators are
phase-volume conserving, time-reversible, and second-order accurate. With these
newly developed particle integrators combined with the higher-order
Godunov hydrodynamic scheme of Athena, this hybrid code
can accurately simulate both hydrodynamics
(e.g. gap opening and RWI) and particle dynamics. 
In this paper, we systematically explore the effects of different mass planets (from far
below the thermal mass to far above the thermal mass) on dust particles
(with particle sizes spanning more than 6 orders of magnitude from
well-coupled limits to decoupled limits) in 2-D and 3-D protoplanetary disks.
In particular, we study whether a very low mass planet can produce prominent
dust gaps, whether particles still settle within the vortex at the gap edge, and particles of what
size can be trapped by the vortex.
 
In Section 2, we introduce the  particle integrators. In Section 3, we describe how
we set up our simulations, together with a discussion on particle settling and drift
in disks. Our results are presented in Section 4. A short discussion is presented in Section 5, and conclusions are
drawn in Section 6. The full description of the newly developed particle integrators,
and various test problems are presented in the Appendix.

\section{Method}
\begin{table*}[ht]
\begin{center}
\caption{Models \label{tab1}}
\begin{tabular}{cccccc}

\tableline\tableline
Case name & Resolution  & box size & Planet mass & R$_{soft}$ & Evolution Time  \\
                      &    R$\times\phi(\times Z)$& $R/H (\times Z/H for 3D)$ &       M$_{T}$  &                     & 2$\pi/\Omega(R=1)$ \\
\tableline
2-D &&&&&\\
\tableline
M0D2 & 400$\times$1024 & [0.5-3] & No &  0.0292 & 400  \\
M02D2 & 400$\times$1024 & [0.5-3] & 0.2  (8 $M_{\oplus}$) &  0.0292 & 400  \\
M10D2 & 400$\times$1024 &  [0.5-3] &1 (0.13 $M_{J}$)&0.05 & 200  \\
M50D2 & 400$\times$1024 &  [0.5-3] & 5 (0.65 $M_{J}$)& 0.0855 & 200  \\
M1JD2 & 400$\times$1024 & [0.5-3] &  1 $M_{J}$ &  0.0855 & 250  \\
M3JD2 & 400$\times$1024 &  [0.5-3] & 3 $M_{J}$&0.0855 & 250  \\
M9JD2 & 800$\times$1024 &  [0.5-5.5] & 9 $M_{J}$& 0.178 & 250  \\
\tableline
3-D &&&&&\\
\tableline
M02D3 & 368$\times$1024$\times$80 & [0.7-3]$\times$[-0.25,0.25] & 0.2 &  0.00855 & 400  \\
M50D3 & 400$\times$1024$\times$160 & [0.5-3]$\times$[-0.5,0.5] &  5 & 0.025 & 150  \\
M50D3H & 800$\times$2048$\times$320 &  [0.5-3]$\times$[-0.5,0.5] & 5 & 0.025 & 100  \\
\tableline
Compare Athena with Fargo &only hydro&&&&\\
\tableline
M50D2A (Athena) & 400$\times$1024 &  [0.5-3] &  5 & 0.0855 & 100  \\
M50D2F (Fargo)& 400$\times$1024 &  [0.5-3] & 5 & 0.0855 & 100  \\
M50D2AH (Athena)& 800$\times$2048 &  [0.5-3] &  5 & 0.0855 & 100  \\
M50D2FH (Fargo)& 800$\times$2048 & [0.5-3] &  5 & 0.0855 & 100  \\
\tableline
\end{tabular}
\end{center}
\end{table*}

The gas disk is simulated with Athena (Stone \etal 2008), a higher-order
Godunov scheme for hydrodynamics and magnetohydrodynamics using the piecewise 
parabolic method (PPM) for
spatial reconstruction (Colella 1984), the corner transport upwind (CTU) method 
for multidimensional integration (Colella 1990), and constrained transport
(CT) to conserve the divergence-free property for magnetic fields
(Gardiner \& Stone 2005, 2008).  To study the interaction between the 
planet and the global disk, 
Athena in cylindrical grids
(Skinner \& Ostriker 2010) has been used. 

We have implemented dust particles in Athena as Lagrangian particles with the
dust-gas coupling drag term, as
\begin{equation}
\frac{d\mathbf{v}_{i}}{dt}=\mathbf{f}_{i}-\frac{\mathbf{v}_{i}-\mathbf{u}}{t_{s}}\,,\label{eq:motion}
\end{equation}
where $\mathbf{v}_{i}$ and $\mathbf{u}$ denote the velocity vectors for particle $i$
and the gas, $\mathbf{f}_{i}$ is the gravitational force experienced by
particle $i$, and $t_{s}$ is the stopping time for this particle due to gas drag. 

Bai \& Stone (2010a)
developed particle integrators in the framework of
the local shearing box approximation in Athena
\footnote{Symplectic
integrators for the shearing box are discussed in Rein \& Tremaine (2011)}. 
Using global cylindrical coordinates, we want to design a particle integrator which 
is symplectic when there is no gas drag, and which naturally couples with Athena's 
gas predictor-corrector CTU scheme when gas drag is needed. The algorithm may 
not need to be more accurate than second-order (the gas evolution is second order accurate)
but should be efficient enough allowing
us to evolve millions or even billions of particles at the same time. Such large
number of particles is needed to obtain a well-sampled spatial distribution dust in the disk. This is crucial for
studying dust feedback to the gas in future. In this study, in order to integrate particle orbits 
in cylindrical grids, we develop
three new particle integrators (1. Leapfrog in Cartesian coordinates, referred as Car SI; 2. Leapfrog in cylindrical coordinates, referred as Cyl SI; 3. Fully implicit integrator
in cylindrical coordinates, referred as Cyl IM). Due to the outstanding performance of the second integrator,
we use the second integrator for all particles except the smallest species (which uses the fully implicit integrator) 
in our planet-disk interaction simulations. 

Implementation of gas drag under Athena's 
 predictor-corrector scheme and interpolation of grid 
quantities to the particle location are described in Bai \& Stone (2010a).
In this work, we choose triangular-shaped cloud (TSC) interpolation scheme.
Orbital advection (Masset 2000, Sorathia \etal 2012) has been applied for the gas.
Unlike Bai \& Stone (2010a) who redesigned the particle integrator considering orbital advection of particles,
we do not modify our particle integrators for orbital advection since we cannot find a simple and proper modification 
which can preserve geometric properties of particle orbits using
cylindrical coordinates.
Instead when calculating the gas drag terms
we simply shift the particles azimuthally to the same reference frame as the gas. 
Orbital advection significantly speeds up our simulations since the numerical
time step is limited by the ratio between the grid size and the deviation of the gas and particle
velocity to the background Keplerian velocity. 
When the planet is present in the disk, the numerical time step is normally limited by eccentric
particles which are excited by the planet. For our 2-D simulations, the time step is normally ($0.01- 0.04)/\Omega(R=1)$, 
while for 3-D simulations,
the time step is normally ($0.005- 0.01)/\Omega(R=1)$.

The numerical algorithms and geometrical properties for these three integrators
are discussed in detail in Appendix A. 
Test problems for particle integrators are presented in Appendix B. 
Most of the tests are designed to study the behavior of an individual particle
in disks, such as drift, settling, perturbations by the planet. They are quite helpful 
for understanding
particle collective behaviors in disks. We encourage the readers to at least skim 
through the test problems.

\section{Model set-up}
We have run global 2-D and 3-D stratified hydrodynamic simulations for up to 400 orbits \footnote{Throughout the paper, if not specified,  the
orbital time means the orbital time at $R$=1 where the planet is placed.} with a minimum resolution of 
8 grids per scale height in each direction. The planet is on a fixed orbit, and have
 a mass of 0.2, 1, or 5 $M_{th}$, equivalent to 8 $M_{\oplus}$, 0.15 $M_{J}$, or 0.65 $M_{J}$ 
(based on Eq. \ref{eq:thermal} with $H/R$=0.05). There are 7 different types of particles from the well-coupled to decoupled limit, each with
one million particles.  The cases with a more massive planet will be discussed 
separately in Section 5.1. 
All simulations
are summarized in Table 1. 

\subsection{Gas component}
Our cylindrical grids span from 0.5 to 3 in the $R$ direction (the planet is at $R=1$) and -$\pi$ to $\pi$ in the $\phi$ direction. For 3-D simulations,  
the $z$ direction extends from -0.5 to 0.5. The grid is uniformly spaced in both $R$ and $z$ directions. The planet is
positioned at $R=1$ and $z=0$ with a Keplerian orbit. 

The disk surface density profile is taken to be
\begin{equation}
\Sigma(R)=\Sigma_{0}(R/R_{0})^{-1}\,, \label{eq:Sigma}
\end{equation}
where $\Sigma_{0}$ and $R_{0}$ are both set to be 1 in code units. For 3-D simulations, the disk also has 
a vertical structure as 
\begin{equation}
\rho(R,z)=\frac{\Sigma(R)}{\sqrt{2 \pi}H(R)}e^{-z^{2}/(2 H(R)^{2})}\,,\label{eq:rho}
\end{equation}
where $H(R)=c_{s}/\Omega(R)$. We use the isothermal equation of state and $c_{s}$ is a constant
in the whole domain, so that $H(R)\propto R^{1.5}$. We set $H(R)/R$ 0.05 at $R=1$. Thus our grid
extends to 10 scale heights vertically from the midplane at $R=1$. The stellar potential for the gaseous disk is set to be 
$\Phi(R,z)=\Omega(R)^{2}z^2/2-1/R$ which is a good approximation for the real potential in a thin disk\footnote{The potential 
is chosen in this form so that in the vertical direction the
disk with density structures as Eq. \ref{eq:rho} is in pressure equilibrium. 
 For particle integrators, we still use $\Phi=1/r$, where $r$ is the spherical radius.}. 
 To avoid a very low density when $z\gg H(R)$, we assume that the potential
is flat beyond $5 H(R)$ and $\rho(z>5 H(R))=\rho(z=5 H(R))$ in the initial condition.
 Both $v_{R}$ and $v_{z}$ are zero initially. $v_{\phi}$ is set
to be balanced by the centrifugal force and the radial pressure gradient.

Fluid variables at both inner and outer boundaries are set to be fixed at their initial values. 
The z direction boundary condition is set 
by extrapolating the density at the last active zone ($\rho_{a}$) to the ghost zones using 
\begin{equation}
\rho_{g}=\rho_{a}\times e^{-(z_{g}^{2}-z_{a}^{2})/(2H^{2})}\,,
\end{equation} 
where $z_{a}$ and $z_{g}$ are the $z$ coordinates of the last active zone and the ghost zones.
 Velocities are copied from the last active zones to the ghost zones. If $v_{z}$ in the last
 active zones will lead to mass inflow
 into the computation domain, $v_{z}$ of the ghost zones is set to be zero.
 We found that these boundary conditions can
ensure a well-established hydrostatic equilibrium disk with velocity fluctuations smaller than 10$^{-6}$ using Athena. 

The planet is introduced after 10 orbits, and its mass is linearly increased to its final value for 10 orbits. To avoid the numerical
divergence of the planet's potential, a fourth order smoothing of the potential has been applied (Dong \etal 2012a). In 2-D simulations,
in order to mimic the planet potential in 3-D, we choose the smoothing length equal to
the planet Hill radius $\sim R_{p}(M_{p}/M_{*})^{1/3}$. A detailed study on the smoothing length is discussed
in M{\"u}ller et al. (2012). In 3-D simulations, the planet potential in 3-D is considered so that 
we can choose a much smaller smoothing length just to avoid numerical divergence (Table 1).

\subsection{Dust component}
In 2-D simulations, we 
distribute particles in a way which leads to the same surface density profile as the gas (Eq. \ref{eq:Sigma}) \footnote{
For passive particles, we can arbitrarily choose the dust surface density. To compare with observations,
we only need to scale the dust density using the realistic dust to gas mass ratio.}.
In detail, to ensure that the particle surface density has a slope of -1, each particle is placed in the disk at 
a constant probability in both R and $\phi$ directions. 
Their velocities
are circular Keplerian velocities so that each particle will maintain the same $R, v_{R}, v_{\phi}$ all the time and
the particle distribution function will not evolve with time.

However, this steady state particle distribution is not trivially achieved in 3-D disks. The vertical motion for
particles is controlled by the Hamiltonian ${\cal H}=(1/2)  v_{z}^{2}+ \Phi(z)$ where $\Phi(z)\sim \Omega^{2}z^{2}/2$,
which describes their vertical oscillations. Thus the dust distribution function will normally evolve with time.
In order to achieve a steady state, we need to derive the particle distribution function from integrals of motion based on Jeans Theorem (Chap 4. of 
Binney \& Tremaine 2008). 
Using the Hamiltonian, which is an integral of motion, we construct the ergodic steady state particle distribution
\begin{equation}
f(v_{z}, z)=e^{-{\cal H}/( \Omega^{2} H_{p}^{2})}=e^{ -v_{z}^{2} /(2 \Omega^{2} H_{p}^{2})-z^{2}/(2 H_{p}^{2})}\,,
\end{equation}
as the initial condition, where $H_{p}$ is the scale height for dust particles. Particles will settle to the disk midplane
on a timescale of $t_{settle}\sim \Omega^{-1}(t_{s}\Omega+(t_{s}\Omega)^{-1})$ (Weidenschilling 1980). 
For particles we studied ($t_{s}\Omega\sim 0.002-200$),
they will settle to the midplane on a timescale of 1000 orbits (see our settling tests in Appendix B).  Since 1000 orbits corresponds to 
1 Myr at 100 AU, 
these particles inside 100 AU normally have already settled to the midplane before the planet formation.
Thus we choose $H_{p}=0.2 H$ as our initial condition for dust particles \footnote{Since the disk is not
turbulent in our simulations, all dust should have already settled to the disk midplane ($H_{p}\sim 0$). 
On the other hand, we want to
study how dust settles in the gap edge vortex later, we thus choose a dust disk with
$H_{p}=0.2 H$, which has significantly
settling but also is resolved by the simulation.}. 

Particles will not only settle but also drift radially in disks (Details on drift tests are presented in Appendix B.).
In our simulations, particles with $t_{s}\Omega\sim 1$ will drift to the central star within several hundreds orbits.
Thus we only run our simulations for at most 400 orbits.

We have evolved seven types/species of particles simultaneously with the gas fluid. We assume that
all these particles are in the Epstein regime, so that
 the dust stopping time (Whipple 1972, Weidenschilling 1977, we use the notation from Takeuchi 
\& Lin 2002) is
\begin{equation}
t_{s}=\frac{ s \rho_{p} }{\rho_{g}v_{T}}\,,\label{eq:ts}
\end{equation}
where $\rho_{g}$ is the gas density, s is the dust particle radius, $\rho_{p}$ is the dust particle density (we 
choose $\rho_{p}$=1 g cm$^{-3}$), v$_{T}$=$\sqrt{8/\pi}c_{s}$, and $c_{s}$ is the gas sound speed. 

For 2-D simulations, we do not have information on how particles are distributed in the $z$ direction, thus
 we have to make the assumption that the majority of dust particles are close to the midplane so that $\rho_{g}$
in Eq. \ref{eq:ts} is the midplane gas density $\rho_{g,mid}=\Sigma_{g}/\sqrt{2\pi} H$. 
In this case, the dust stopping time can be written as
\begin{equation}
t_{s}=\frac{\pi s \rho_{p} }{2\Sigma_{g}\Omega}\,.\label{eq:ts2}
\end{equation}
The assumption that  the majority of dust particles are close to the midplane will be justified later in Section 3.3. However,
we want to caution that, in 2-D simulations, there is an inconsistency in using the 2-D pressure 
to calculate the particle drift speed, which is discussed in detail at the end of Appendix B. 

In a dimensionless form the stopping time for both 2-D and 3-D cases can be written as 
\begin{equation}
T_{s}=t_{s}\Omega\,,
\end{equation}
which is normally referred as the Stokes number (St).

To make our results general, we do not specify the length and mass unit in our simulations. 
In our simulations, we have seven particle types and 
each particle type has the same $s$ in Eqs. \ref{eq:ts} and \ref{eq:ts2}.
With our chosen $s$, $T_{s}$ for different particle types at $R=1$ in the initial
condition are shown in Table ,2 and the radial profile of $T_{s}$ at the disk midplane are shown in the
lower right panel of the Figure in Appendix B, 
\begin{equation}
T_{s}=\frac{\pi s \rho_{p} }{2\Sigma_{g}}\propto R\,. \label{eq:TsR}
\end{equation}
Note that for every particle, $t_{s}$ will evolve with time since
$\rho_{g}$ or $\Sigma_{g}$ evolves with time. 

A given $T_{s}$ corresponds to some particle
size if we assign the realistic disk surface density and the planet position to our simulations.
The relationship
between $T_{s}$ and disks' physical units can be derived by 
\begin{equation}
T_{s}=1.55\times10^{-3}\frac{\rho_{p}}{1 {\rm g \, cm^{-3}}}\frac{s}{1\, {\rm mm}}\frac{100\, {\rm g \,cm^{-2}}}{\Sigma_{g}}\,.\label{eq:ts3}
\end{equation}
For example, if we are studying $\alpha$ disk similarity solution with $\alpha=0.01$, $\dot{M}=10^{-8}\msunyr$,
and $ T = 221 (R/{\rm AU})^{-1/2}$~K, the disk surface density is $\Sigma_{g}=178 (R/AU)^{-1}$ g cm$^{-2}$. Then
the smallest species with $T_{s}=0.001765$ at $R=1$ corresponds to 
0.4 mm particles at 5 AU and 0.1 mm particles at 20 AU.  However, with a realistic disk surface density,
some particle types are in the Stokes regime and our Epstein regime assumption breaks down. Those particle types can
only be considered as a numerical experiment to explore the effect of a large stopping time on particle distribution. 
The corresponding particle sizes for different realistic disk structures are summarized in Table 2.

\begin{table}[ht]
\begin{center}
\caption{Corresponding particle sizes \label{tab1}}
\begin{tabular}{lccccc}
 
\tableline\tableline
Par.  & $t_{s}\Omega$ & Par. size & Par. size & Par. size  &  Par. size \\
         type               &   at R=1    &   at 5 AU in  a  &   at 5 AU in &  at  20 AU in      &   at  20 AU in            \\
                        & initially & MMSN disk\tablenotemark{1}  &  $\alpha$ disk A\tablenotemark{2} & $\alpha$ disk A & $\alpha$ disk B\tablenotemark{3} \\
\tableline
a &   0.001765   &  1.7 mm & 0.4 mm & 0.1 mm & 1 mm \\
b &   0.01765   & 1.7 cm & 4 mm &  1 mm  &  1 cm  \\
c &   0.1765 & 17 cm &  4 cm & 1 cm  & 10 cm \\
d &   1.765  &  Stokes\tablenotemark{4} & 40 cm & 10 cm   & 1 m\\
e  &   17.65 &  Stokes &  4 m  & 1 m & 10 m \\
f  &   176.5 &  Stokes &  Stokes  & 10 m & Stokes  \\
g &   $\infty$ &  - &  -  & - & -  \\
\tableline
\end{tabular}
\tablenotetext{1}{ A MMSN disk has $\Sigma_{g}=1700 (R/AU)^{-1.5}$ g cm$^{-2}$.}
\tablenotetext{2}{ A $\alpha$ disk A has $\Sigma_{g}=178 (R/AU)^{-1}$ g cm$^{-2}$, which is the surface density of a constant $\alpha=0.01$ accretion disk with
$\dot{M}=10^{-8}\msunyr$,
and $ T = 221 (R/{\rm AU})^{-1/2}$~K }
\tablenotetext{3}{ A $\alpha$ disk B has $\Sigma_{g}=1780 (R/AU)^{-1}$ g cm$^{-2}$, which is the surface density of a constant $\alpha=0.001$ accretion disk with
$\dot{M}=10^{-8}\msunyr$,
and $ T = 221 (R/{\rm AU})^{-1/2}$~K }
\tablenotetext{4}{ Particles in Stokes regime, see text.} 
\end{center}
\end{table}

If the reader only wants to pick one particle size for one particle type in our simulations  to 
picture how different sized particles distribute in a typical protoplanetary disk, 
one can consider, a planet at 20 AU, Par. a as 0.1 mm, Par. b as 1 mm, Par.c as 1 cm,
and so on in a typical protoplanetary disk with $\Sigma_{g}=178 (R/AU)^{-1}$ g cm$^{-2}$.

\section{Results}
In our subsequent 2-D and 3-D simulations with planets of different masses, we first present the results for gas
density distribution (Section 4.1). Then we describe spatial 
distribution of particles which couple with the gas relatively well (Section 4.2).
Finally, we show the results for particles which  are almost completely decoupled from the gas (Section 4.3).  

\begin{figure*}[ht]
\centering
\includegraphics[width=0.78\textwidth]{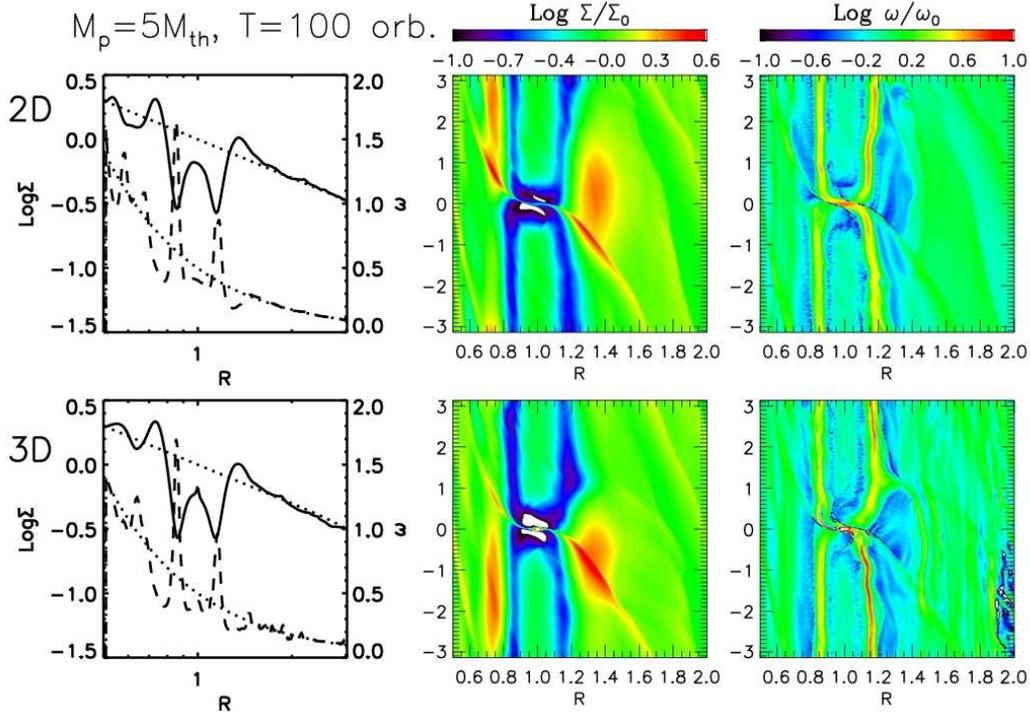}
\vspace{-0.3 cm}
\caption{The gas surface density and vorticity 
at 100 orbits for both 2-D (upper panels) and 
3-D (lower panels) simulations with
a 5 $M_{th}$ (0.65 $M_{J}$) planet. Left panels: the azimuthally averaged disk surface density (solid curves)
and vorticity (dashed curves. For the 3-D case, the vorticity is derived at the midplane).  The dotted curves
are the density and vorticity profiles from the initial condition. Middle
panels: the disk surface density in the $R-\phi$ plane normalized to the initial surface density. Right panels: the vorticity in the $R-\phi$
plane normalized to the initial vorticity (for the 3-D case, the vorticity is derived at the midplane). 
Throughout the paper, we will plot our 2-D and 3-D cylindrical global
simulations in the $R-\phi$ plane to better demonstrate the detailed flow structure.}
\label{fig:fig3p2}
\end{figure*}

\subsection{Gaseous Disks}
Massive planets open gaps in the gas quickly, while very low mass planets can also induce gaps far away from the planet
 by nonlinear wave steepening.

\subsubsection{Massive Planets}

\begin{figure*}[ht!]
\centering
\includegraphics[width=0.78\textwidth]{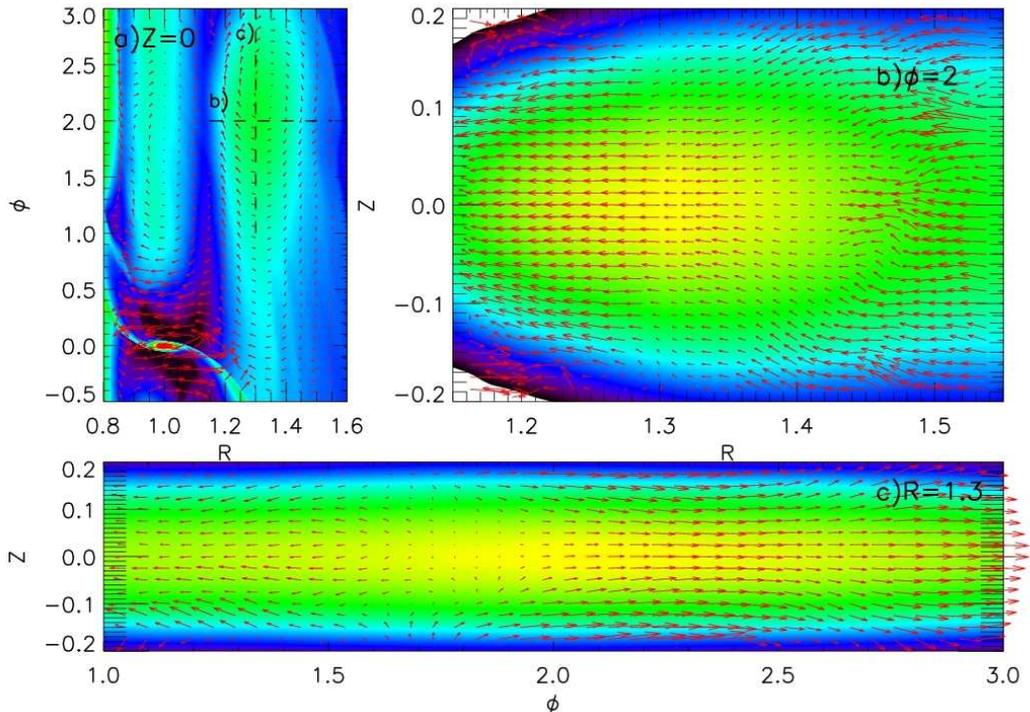} 
\vspace{-0.3 cm}
\caption{The 3-D vortex structure from the
M50D3 case at 90 orbits. Velocity vectors are overplotted onto the density color contours (Keplerian velocities
are subtracted from $v_{\phi}$). The upper left panel shows the density contours at the disk midplane in the $R-\phi$ plane.
The vortex is clearly anti-cyclone with a higher density at the center. Two perpendicular
cuts across the vortex are shown in the upper right and lower panels.
} \label{fig:fig12}
\end{figure*}

\begin{figure*}[ht]
\centering
\includegraphics[width=0.80\textwidth]{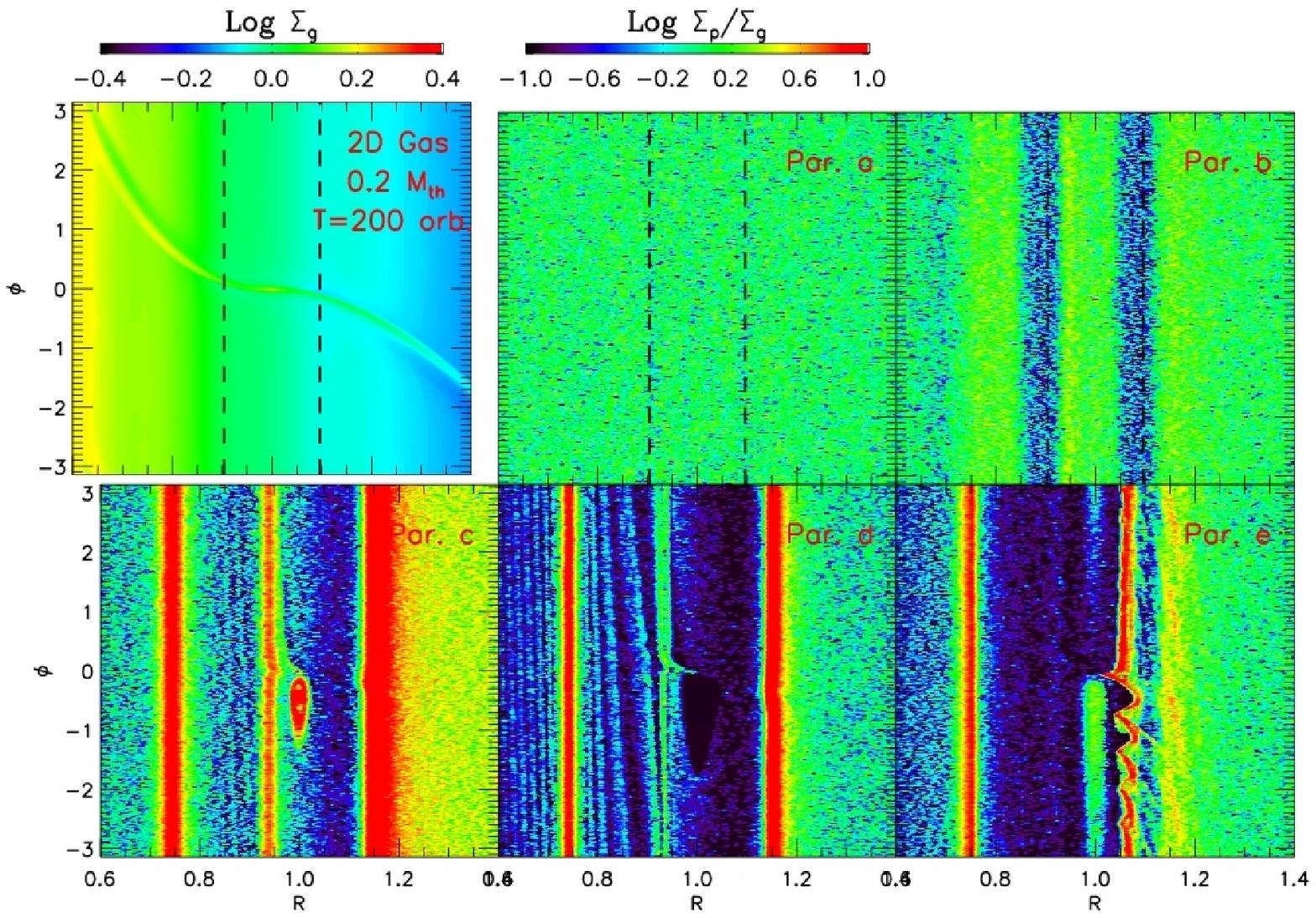} 
\vspace{-0.3 cm}
\caption{
The disk surface density from the M02D2 case at 200 orbits.
The upper left panel shows the surface density of the gas disk.
The dashed lines label where the spiral density waves shock and gap opening occurs.
However, the gaps in the gas are so shallow that they are not visible in this plot (the gaps
are more visible in Figure \ref{fig:fig4p1}). 
The other five panels show the ratio between the dust surface density
and the gas surface density at 200 orbits.
The smallest particles (Par. a) couple with the gas so well that the ratio is one. The shallow gaseous
gaps start to concentrate Par. b at the gap edges. 
This dust concentration 
at the gap edges and the co-orbital region are apparent for
Par. b to Par. d.
Par. e starts to decouple from the gas and the eccentricity of particles is driven up by the planet when they
move close to the planet. 
 After several orbits the particle eccentricity is gradually damped.  This decreasing amplitude of the ripple has
 analogy with the ripple excited by Pan in the Encke gap of Saturn's rings.
} \label{fig:fig4}
\end{figure*}

Gaps in the gas are quickly induced by 1 $M_{th}$ and 5 $M_{th}$ planets in disks. More interestingly,
with gaps getting deeper and gap edges becoming sharper, the gap edges become Rossby wave unstable and start to develop vortices. 
Figure \ref{fig:fig3p2} shows the development of vortices in both 2-D and 3-D disks. In the leftmost panels,
the azimuthally averaged density and vorticity ($\omega_{z}=(\nabla\times \mathbf{v})_{z}$) are overplotted. The vorticity peaks 
where the gap is deepest. The $R-\phi$ 2-D distribution for density and vorticity
are shown in the middle and rightmost panels of Figure \ref{fig:fig3p2}, where the high vorticity
region outlines the horseshoe orbits. Vortensity ($\omega/\Sigma$) is known to be conserved in a 2-D barotropic flow. Only shocks can generate/destroy vortensity. 
Strong shocks are excited around the planet when the planet mass is
larger than $M_{th}$. These shocks can change the vortensity in two ways. 
First they can add vortensity to the fluid in horseshoe orbits when the flow is passing the planet, 
and these high vortensity gas circulates in the horseshoe region
(Lin \& Papaloizou 2010). Second, the shock will push the flow away from the planet and lead to
gap opening, which causes
 a vortensity minimum at the gap edge (Li \etal 2009). 
Since vortensity is not 
conserved in 3-D disks, the above explanations cannot be applied to 3-D disks. However
the great similarity between the upper (2-D cases) and lower (3-D cases) panels of Figure \ref{fig:fig3p2} suggests
that, at the midplane of the 3-D disk, the disk behaves approximately two dimensionally.

Although the peaks of vortensity just outside the horseshoe region are prominent, 
RWI is not operating there. 
As Lin \& Papaloizou (2010) pointed out,
the gap edge (outside the horseshoe region) where the vortensity is at the minimum  is
Rossby wave unstable. When the RWI first develops, it produces 3-4 vortices at the gap edge,
which was analytically understood by linear analysis of Lin \& Papaloizou (2010) when $H/R\sim0.05$.
Eventually, these vortices merge into a single vortex. The vortices have smaller
vorticity than the background flow, and thus are anti-cyclones with higher density at the center.

A more detailed analysis of the 3-D velocity field of the vortex in Figure \ref{fig:fig12}
also suggests that the vortex is basically two dimensional. 
The slices
across the vortex at the $R-\phi$ plane, $R-z$ plane, and $\phi-z$ plane are shown
in Figure \ref{fig:fig12}. The slices in Panel b) and c) are slightly
offset the center of the vortex to show the rotation structure of the vortex. Within one scale
height of the disk midplane ($|z|<0.05$), the vortex has very little
vertical motions. 
Only beyond two scale heights ($z>0.1$), the flow 
has a vertical velocity towards
the midplane when it is behind the center of the 
vortex in $\phi$ direction, and has a vertical velocity leaving
the midplane when it is  in front of the center of the vortex. 
The 2-D nature of the vortex in our simulations
is consistent with Lin (2012a).
Generally, the 2-D nature
of these vortices suggests that particles will not be disturbed in the vortex, and strong
particle settling and concentration are expected at the midplane
as shown in Section 4.2.

The 3-D vortex at the gap edge does not show turbulent substructures,
 which implies the absence of the elliptical instability in 3-D vortices. We think there may be two reasons
 for this absence. First, the growth rate of the elliptical instability is very small (Lesur \& Papaloizou 2009). For our elongated gap edge vortex with an aspect
ratio $\sim$ 10, it takes hundreds of orbits for the instability to grow, while our simulations only run for 150 orbits. Second, the elliptical instability
requires closed streamlines. However, the streamlines in 3-D votices are not necessarily closed (Zhu \& Stone, in prep.). Moreover, the vortex is constantly
perturbed by the planetary wake, leading to a dynamically evolving vortex with unclosed streamlines.

\subsubsection{Low Mass Planets}

\begin{figure}
\centering
\includegraphics[width=0.5\textwidth]{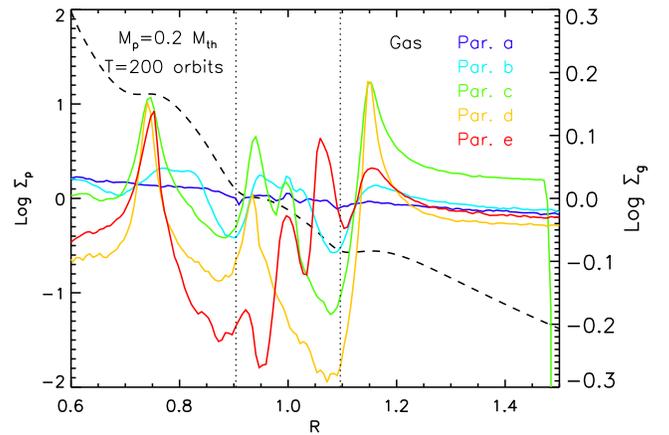} 
\vspace{-0.3 cm}
\caption{
The azimuthally
averaged surface density for the gas (the black dashed curve with the right $y$ axis) 
and different particle types (colored curves with the left $y$ axis) for
M02D2. Two shallow gaps in the gas are opened by the planet at the spiral wake shocking distances,
which are labeled as the dotted vertical lines. At the gap edges, for Par.  c and d, the ratio between the particle and gas 
surface density
can be larger than a factor of 10. Particles are also concentrated inside the horseshoe region.
} \label{fig:fig4p1}
\end{figure}

Contrary to the conventional wisdom (Lin \& Papaloizou 1993), it has been demonstrated recently that even the
low mass planet can open gaps by nonlinear  steepening of the planetary wake as long as the planet 
migration is slow (Goodman \& Rafikov 2001, Rafikov 2002b). Linear waves excited by a low mass planet ($M_{p}<M_{th}$) steepen
to shocks after they propagate a distance
\begin{equation}
|x_{sh}|\approx 0.93 \left(\frac{\gamma+1}{12/5}\frac{M_{p}}{M_{th}}\right)^{-2/5} H\,.\label{eq:eqshock}
\end{equation} 
where $\gamma$ is the adiabatic index. Shocks efficiently 
transfer the wave
angular momentum to the background disk flow,
and two gaps gradually open
beyond $|x_{sh}|$ on each side of the planet. 
The nonlinear wave steepening theory has been verified by numerous
simulations (Muto \etal 2010, Dong \etal 2011b, Duffell 
\& MacFadyen 2012, Zhu \etal 2013).
In our M02D2 case where we have a $0.2
M_{th}$ mass planet in an isothermal disk ($\gamma$=1), the shocking
distance is $|x_{sh}|\sim 0.95 H$.  The two radii where planetary wakes
start to shock are labeled as the vertical dashed lines in Figure \ref{fig:fig4} and dotted
lines in Figure \ref{fig:fig4p1}. Unfortunately, the gap opening timescale 
for a  0.2 $M_{th}$ mass planet is $\sim$1000 orbits (Zhu \etal 2013). 
For only $\sim$ 200 orbits in our simulations M02D2, 
the two gaps in the gas are so shallow that they are not visible in the upper leftmost panel of
Figure \ref{fig:fig4}, and are only marginally visible in Figure \ref{fig:fig4p1}  as the dashed curve. However,
these shallow gaps have significant effects on dust particles as shown in Section 4.2.

\subsection{Well/Moderately Coupled Particles in Disks}
\begin{figure*}[ht!]
\centering
\includegraphics[width=0.8\textwidth]{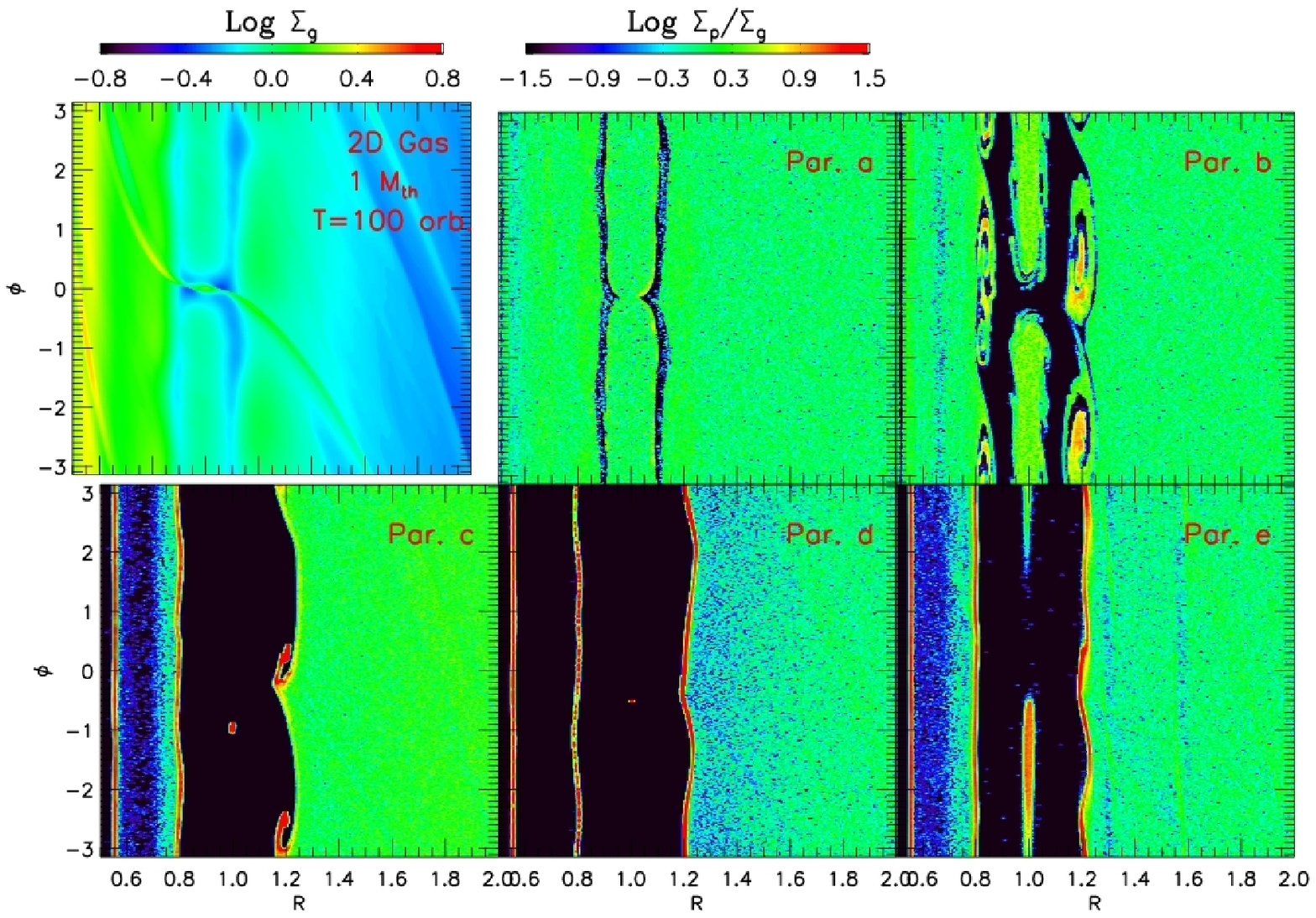} 
\vspace{-0.4 cm}
\caption{
Similar to Figure \ref{fig:fig4} but with a 1 $M_{th}$ (0.13 $M_{J}$) planet in the disk (M10D2).
Gap edges become unstable and vortices start to develop. Vortices
can concentrate particles. Par. b reveals a detailed flow structure in the vortex. Particles
with $T_{s}\sim 1$ (eg. Par. c and d) have the highest concentration in the vortex.
For Par. c to Par. e, particles in the horseshoe
region seem to be only present  at the side of the L5 Lagrangian point.} \label{fig:fig5}

\centering
\includegraphics[width=0.8\textwidth]{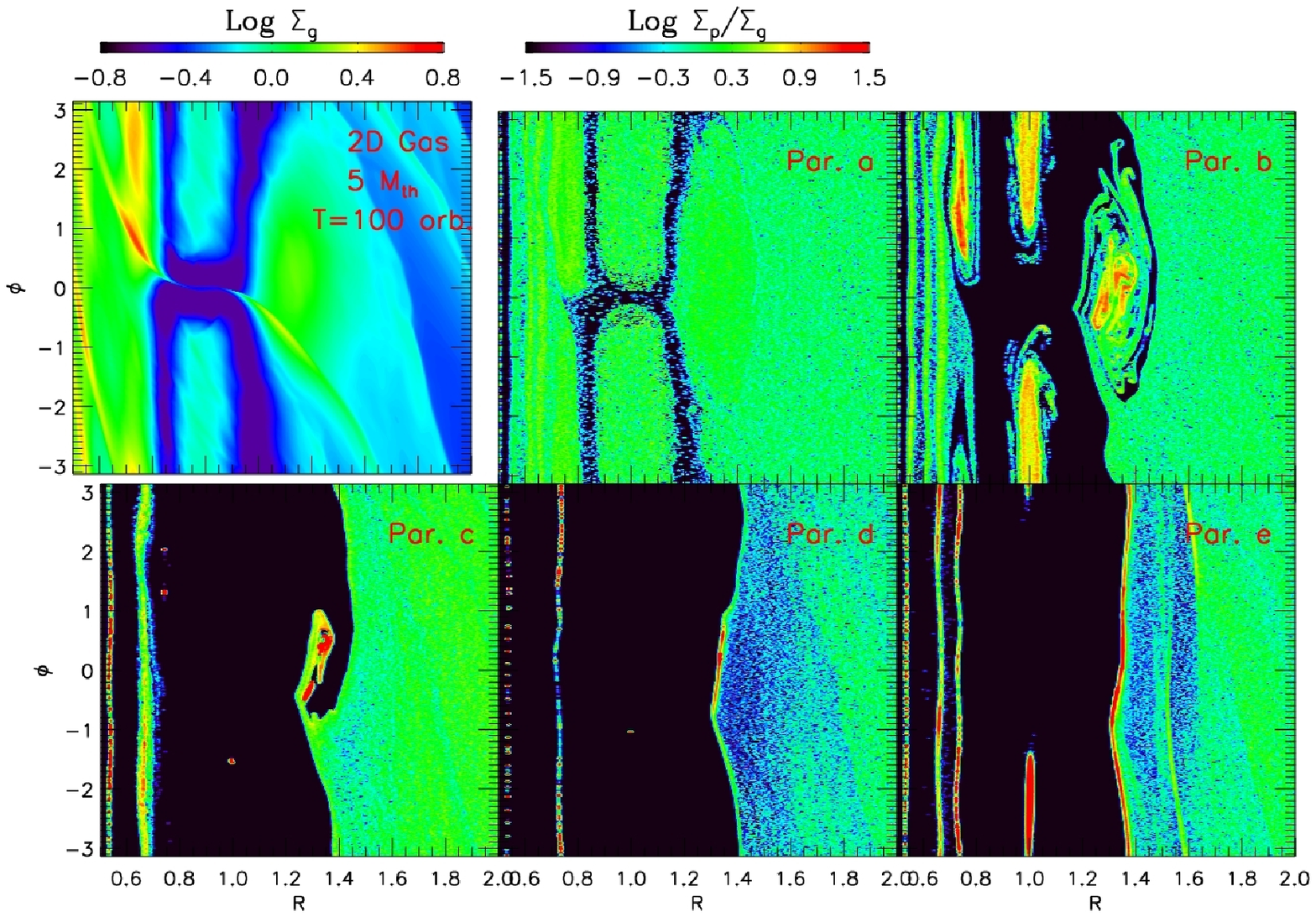} 
\vspace{-0.4 cm}
\caption{
Similar to Figure \ref{fig:fig4} but with a 5 $M_{th}$ (0.65 $M_{J}$) planet in the disk (M50D2).
Gaps are deeper than those in the M10D2 case. The vortices have merged
into one large vortex. The dust concentration ratio can reach to more than 100 within
the vortex for Par. b and Par. c. For Par. e, gaps are opened at $R\sim 1.6$, which is the 2:1 mean motion resonance
with the planet. 
} \label{fig:fig6}
\end{figure*}

In this subsection, we will focus on particles with $T_{s}\le 10$ (Particle types a-e). 
The dust surface densities in 2D simulations (M02D2, M10D2, M50D2) are shown
in Figure \ref{fig:fig4} to \ref{fig:fig6p1}, while dust distributions in 3D simulations (M02D3, M50D3)
are shown in Figure \ref{fig:fig8} to \ref{fig:fig11}. 

\subsubsection{2-D Disks}
Figures \ref{fig:fig4}-\ref{fig:fig6} show the gas and
 dust distributions
 from the 2D simulations M02D2, M10D2, M50D2 respectively. 
 The color contours in Par. a to Par. e 
 panels display the 
surface density ratio between each dust type and the gas. 
 We refer to this dust-to-gas mass ratio as 
 the dust concentration ratio. Since each particle
 species is set to have the same surface density as the gas initially,
this dust concentration ratio
is one in the initial condition. Particles
in our simulations
are passive particles without feedback to the gas, so this initial dust-to-gas mass
ratio can be scaled to any realistic dust-to-gas mass ratio
in a protoplanetary disk. For example, assume that a realistic protoplanetary disk
can be represented by our M50D2 case, but
the real mass
ratio between 1 mm dust particles and the gas is 0.001 in most parts of the
disk. The fact that the dust concentration ratio for 1 mm particles is 100 in the vortex based on our simulations
means that the real mass ratio between  1 mm dust and gas in the vortex of this disk is $0.001\times 100=0.1$.

Even with a 0.2 $M_{th}$ (or 8 $M_{\oplus}$) planet in the disk, 
Figure \ref{fig:fig4} shows that particles of different sizes have very different distributions in the disk.
Par. a couples
with the gas so well that the dust
concentration ratio is unity everywhere.
For Par. b, the very shallow gaps in the gas at the shocking distance 
(dashed vertical lines) start to
affect particle concentration (also in Figure \ref{fig:fig4p1}). 
Particles are concentrated at gap edges and the planet corotation region. The
effect from the shallow gaps in the gas on dust particles becomes more prominent for Par. c and Par. d, which have the fastest drift speed since their
stopping times are close to one. Considering Par. c and Par. d correspond to 1mm and 1cm dusts at 20 AU
in $\Sigma_{g}=178 (R/AU)^{-1}$ g cm$^{-2}$ disks (Table 2). These gaps may be visible from mm and cm observations. For Par. e which
has $T_{s}\sim 10$ at $R=1$, the eccentricity excitation for particles passing close to the planet becomes visible. 
In the Par. e panel, the ripples outside the planetary orbit are associated with epicyclic motion of dust particles. The 
eccentricity is pumped up when the particles encounter the planet and gradually damps
after several orbits. The eccentric particles are only visible
outside the planetary orbit, because inside the planetary orbit particles
drift inwards and do not experience close encounters with the planet.

The azimuthally averaged density profiles for the gas and dust in M02D2 at 200 orbits 
are shown in Figure \ref{fig:fig4p1}. Two shallow gaps in the gas (dashed curve) 
change particle radial drift speed and 
thus cause particle pile up at the gap edges. Par. c and d have significant
enhancement at the gap edge $R\sim 1.2$ since they have $T_{s}\sim 1$. 
Smaller particles (Par. a) have good coupling to the gas and drift slowest,
so that they obey the same surface density distribution as the gas. Bigger particles (Par. e) start to decouple
from the gas, and extend closer to the planet.

With a 1 $M_{th}$ (0.13 $M_{J}$)
planet in the disk, Figure \ref{fig:fig5} shows the gas and dust distribution from M10D2.
 The gap in the gas is pronounced, which affects all types of particles. Particles at the
outer disk are trapped at the outer gap edge, and cannot move to the inner disk.
Particles which are initially at the inner disk drift to the central star, leaving a large 
particle gap/cavity.
The inner gap edge can also trap some particles. The most prominent features compared with
M02D2 are
two vortices
at each edge of the gap, which concentrate particles significantly. Except
Par. a,  Par. b to Par. e all have some degree of dust 
concentration within the vortices.
Par. b reveals a detailed flow structure since it couples with the gas at a short stopping time
$T_{s}\sim 0.02$ and can trace the small scale features of the gas vortices. 
Par. c has higher dust concentration than Par. b;
the dust concentration ratio can reach 100 at the center of the vortices.  
Par. d and e have stopping times $T_{s}>1$ so that they decouple from the gas
and cannot trace vortex
structures having dynamical timescales smaller than the orbital timescale. Thus Par. d and e.
are almost featureless. Nevertheless the vortices still have secular effects on these particles (Section 4.2.2)
leading to dust concentration within the vortices. 
Since the disk is inviscid, the gaps continue to 
deepen with time and two vortices will eventually merge into one. 

\begin{figure}[ht]
\includegraphics[width=0.5\textwidth]{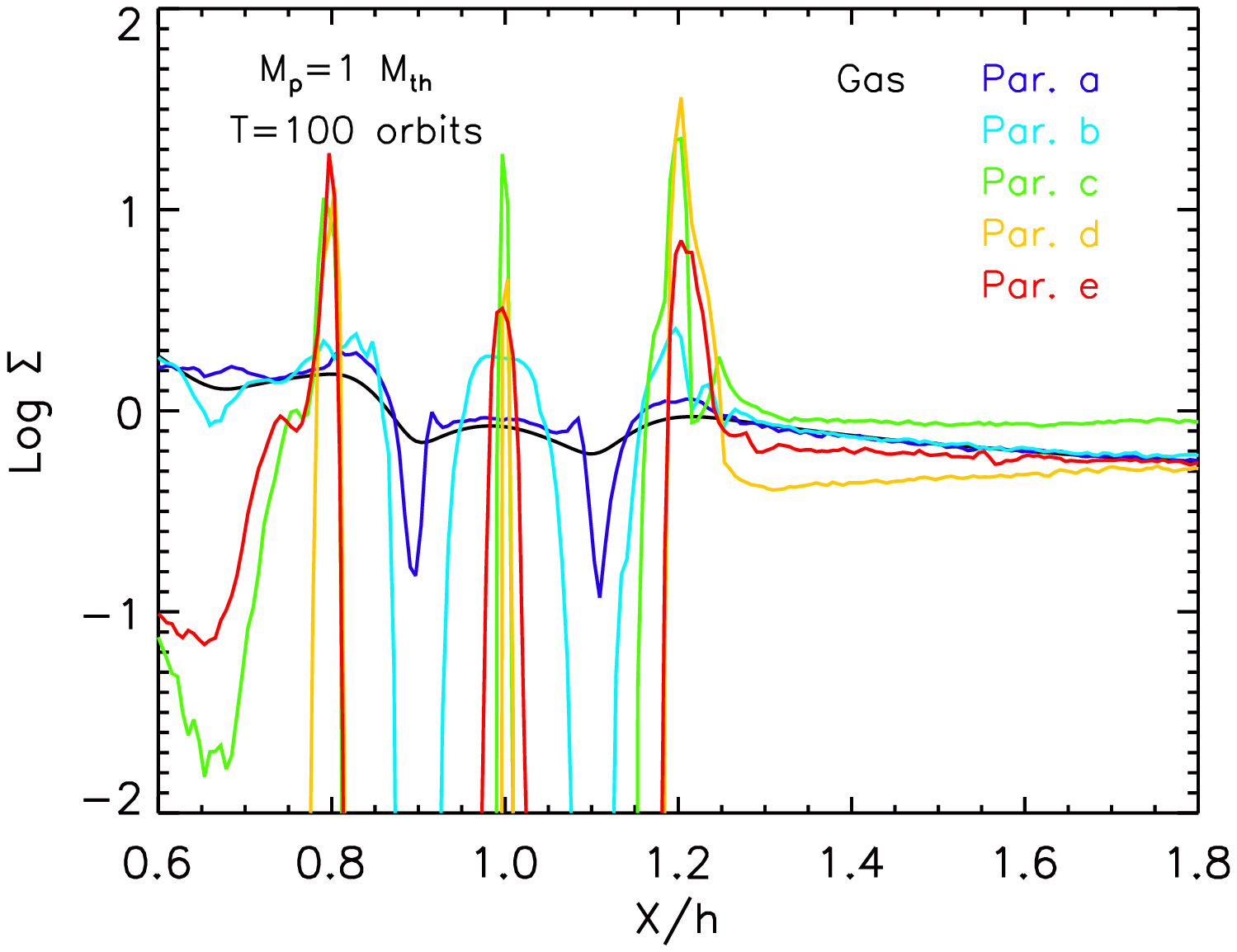} 
\vspace{-0.3 cm}
\caption{
The azimuthally
averaged disk surface density for the gas (black curves) and different particles (colored curves) for
M10D2. } \label{fig:fig5p1}
\end{figure}

Another noticeable effect is that,
for Par. c to Par. e, particles in the horseshoe region seem to be  only present on the same side as 
the L5 point, which suggests that, when there is gas drag, particles on the L5 side of
the horseshoe region are more stable than the particles on the L4 side. The difference in stability
properties of L4 and L5 Lagrange points in presence of a drag force has been
previously addressed by Murray (1994) and Lyra \etal (2009).

The azimuthally averaged density profiles from M10D2 are shown in Figure \ref{fig:fig5p1}.
The factor of 2 density dip in the gap can lead to more than a factor of 10 increase in the dust surface
density at the gap edge within only 100 orbits. The horseshoe region can also trap dust
particles significantly.

\begin{figure}[ht]
\includegraphics[width=0.5\textwidth]{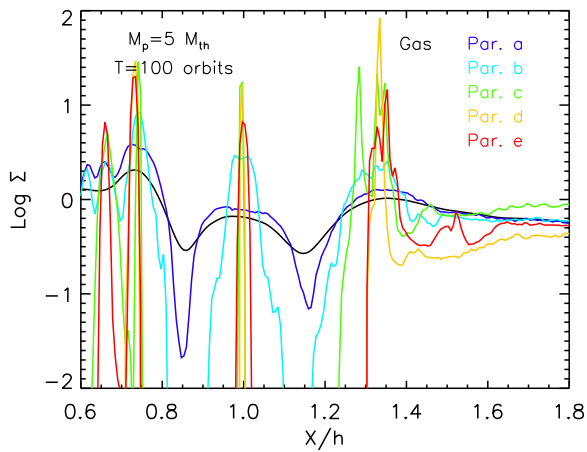} 
\vspace{-0.3 cm}
\caption{
The azimuthally
averaged disk surface density for the gas (black curves) and different particles (colored curves) for
M50D2. } \label{fig:fig6p1}
\end{figure}

With a 5 $M_{th}$ (0.65 $M_{J}$) planet in the disk, 
Figure \ref{fig:fig6}  shows the surface density from M50D2.
The gap in the gas is so deep that vortices quickly 
develop and later merge into one big
vortex.  Again Par. b to Par. e (equivalent to particles with
$T_{s}>0.02$ and $T_{s}<20$) are trapped into the vortex, and particles
at the L5 side of the horseshoe region seem to be more stable than the L4 side for large particles (Par.c to
Par. e).  For Par. e, we also noticed that gaps start to be opened at mean motion
resonances (2:1 mean motion resonance at $R=1.6$), which will be discussed in
detail in Section 4.3. 
The azimuthally averaged density is shown in Figure \ref{fig:fig6p1}. 
Again, particles are significantly trapped at the gap edge and inside the horseshoe region.

\subsubsection{3-D Disks}

\begin{figure*}[ht] 
\centering
\includegraphics[width=0.8\textwidth]{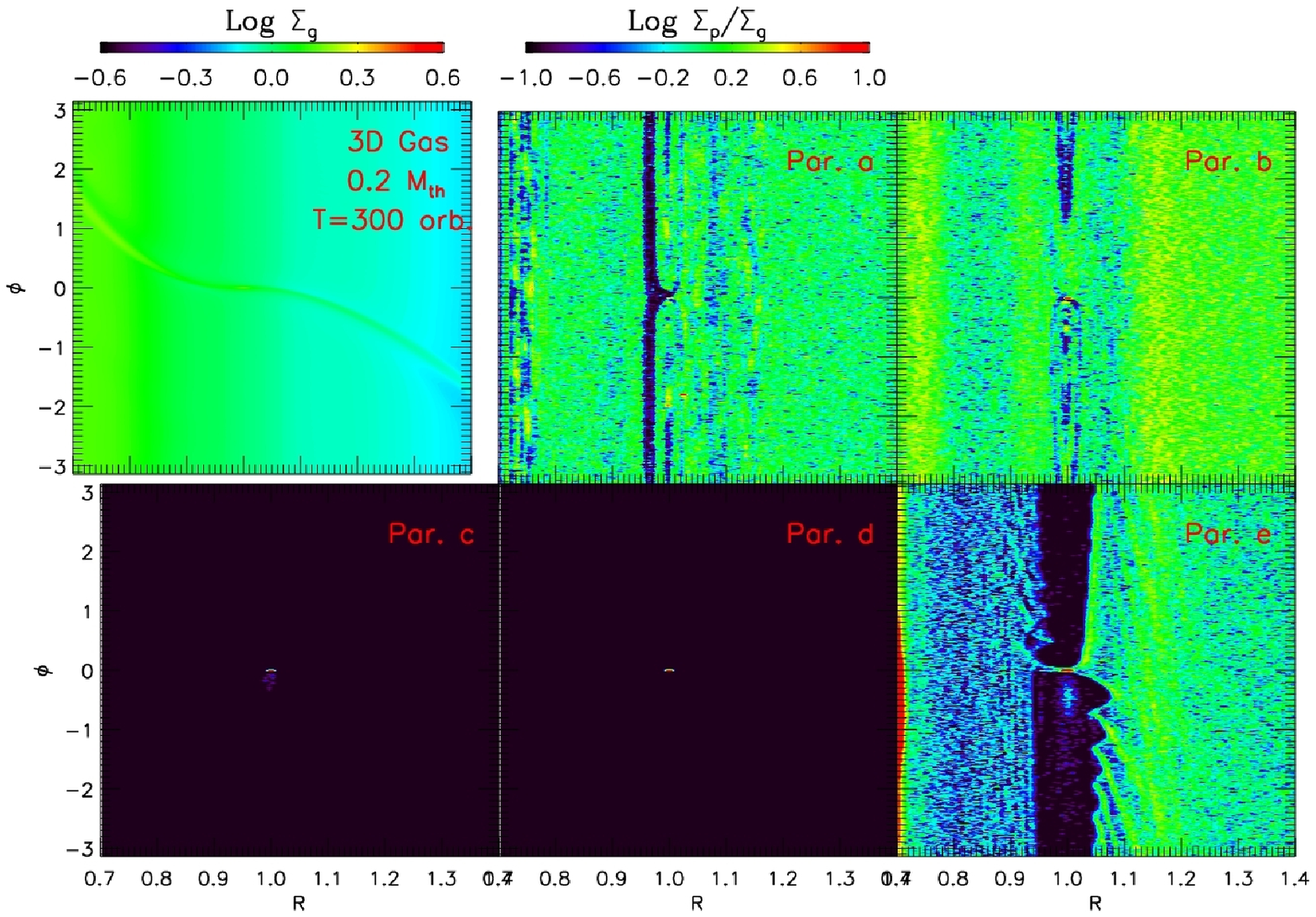} 
\vspace{-0.4 cm}
\caption{
Similar to Figure \ref{fig:fig4} but from the 3-D simulation (M02D3).
3-D and 2-D cases are similar except that 1) the gaps in the gas in 3-D disks
are slightly shallower so that dust concentration is not as prominent as in 2-D; 2) particles
drift faster in 3D disks (Appendix B) so that all Par. c and Par. d have drifted to the central star
before the planet opens a gap. 
} \label{fig:fig8}

\centering
\includegraphics[width=0.8\textwidth]{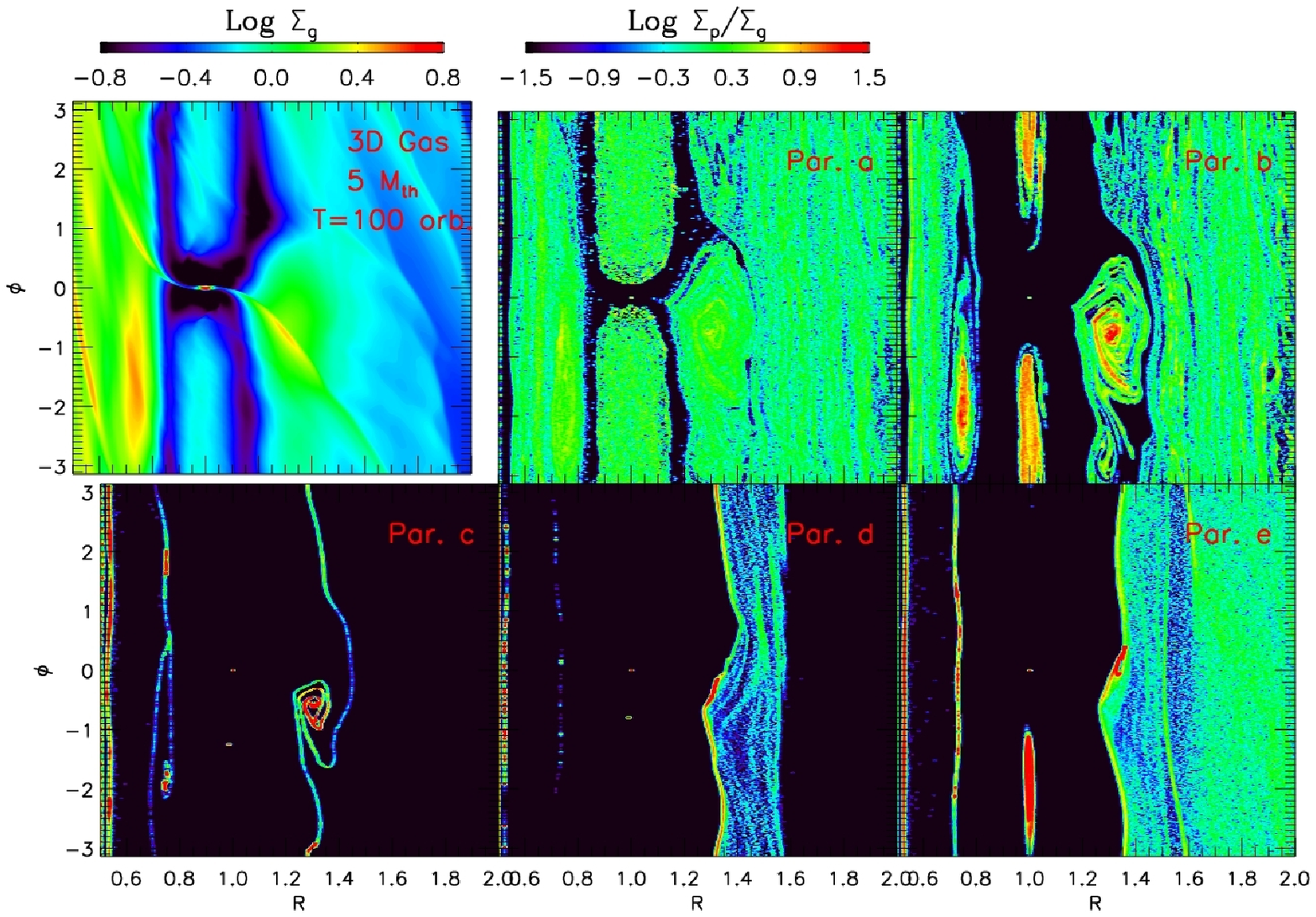} 
\vspace{-0.4 cm}
\caption{
Similar to Figure \ref{fig:fig6} but from the 3-D simulation (M50D3).
Compared with Figure \ref{fig:fig6}, the 3-D case is similar to the 2-D case.
} \label{fig:fig9}
\end{figure*}
The dust surface densities for 3-D simulations (M02D3 and M50D3) 
are shown in Figure \ref{fig:fig8} and Figure \ref{fig:fig9}. Generally, particle surface densities 
in 3-D
disks are very similar to those in 2-D disks by comparing Figure \ref{fig:fig8} \& \ref{fig:fig9} with Figure \ref{fig:fig4} \& \ref{fig:fig6}.
 But there are several noticeable differences
due to the 3-D structure of the disk.

First, particles drift faster at
the midplane of a 3-D disk than in a 2-D disk as discussed in Appendix B. 
In Figure \ref{fig:fig9}, Par. c and Par. d
have all drifted inward of $R\sim1.6$ at 100 orbits in 3-D disks, while there are still particles
beyond $R\sim 2$ at the same time in 2-D disks as shown in Figure \ref{fig:fig6}.
In Figure \ref{fig:fig8},
Par. c and Par. d have all drifted
to the central star within 300 orbits. 

Second, with a 0.2 $M_{th}$ (8 $M_{\oplus}$) planet in the disk, 
the particle gap for Par. b in the 3-D disk (Figure \ref{fig:fig8}) is less prominent than
that in the 2-D disk (Figure \ref{fig:fig4}). This is because the torque exerted by a planet
is 2.5 times stronger in 2-D disks than that in 3-D disks (Tanaka \etal 2002). Thus gaps in the gas are
more difficult to open in 3-D disks than in 2-D disks, and 
shallower gaps in 3-D disks have less effect on particle concentration.

Third, there is particle concentration at the planet's position in 3-D disks. 
By comparing Figure \ref{fig:fig4} and Figure \ref{fig:fig8}, or comparing Figure \ref{fig:fig6}
and Figure \ref{fig:fig9}, we can see that, for Par. c and Par. d, there are no particles at
the planet position ($R=1$, $\phi=0$) in 2-D simulations, while the dust concentration ratios are quite high there 
in 3-D simulations. This is because, in 3-D simulations, a small 
smoothing length for the planetary potential leads to a centrally peaked dense core
around the planet. Such a core has a sharp 
pressure gradient towards the center and
thus can effectively concentrate particles.  
While in 2-D simulations, in order to mimic the planet potential in 3-D, 
a large smoothing  length which is around the planet's Hill radius or the disk scale height is normally applied, leading to
 a smoother core around the planet, which is less efficient at trapping particles.

\begin{figure*}[ht] 
\centering
\includegraphics[width=0.8\textwidth]{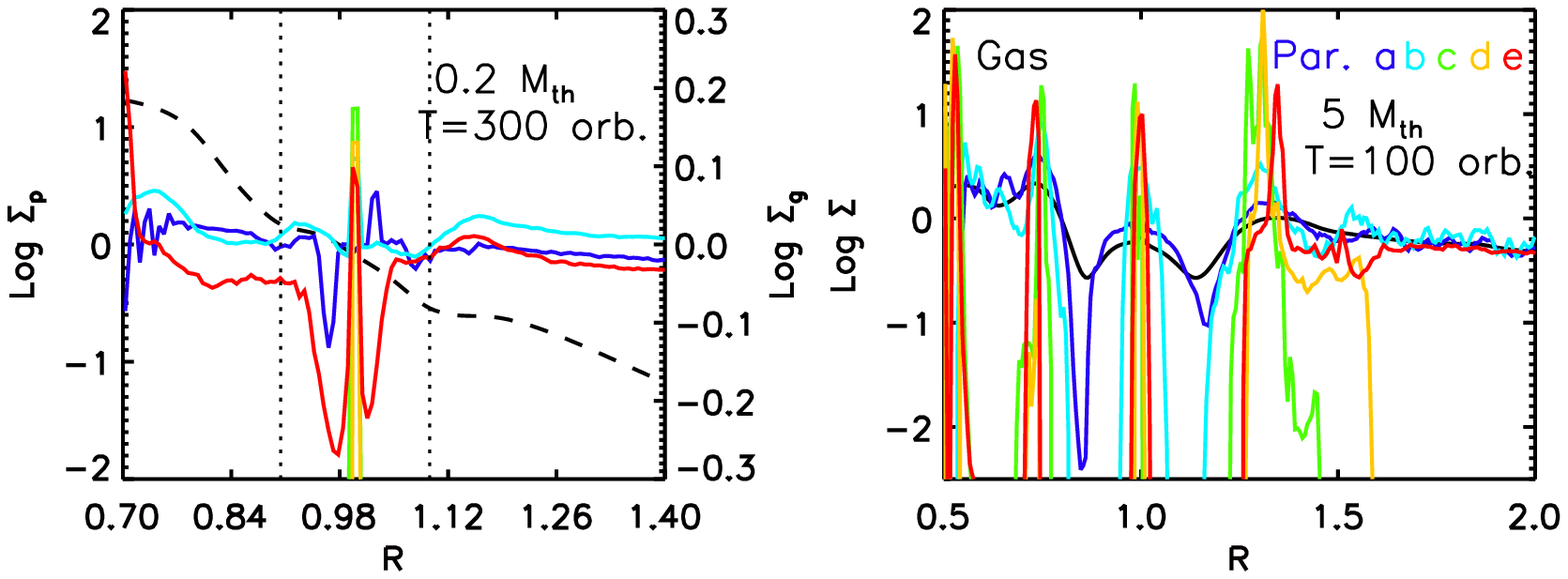} 
\vspace{-4.8 cm}
\caption{Similar to Figure \ref{fig:fig4p1} and \ref{fig:fig6p1}.
The azimuthally
averaged disk surface density for the gas (black curves) and different particles (colored curves) in 3-D cases
M02D3, and M50D3.  } \label{fig:fig10}
\end{figure*}

The azimuthally averaged surface density is shown in Figure \ref{fig:fig10}. By comparing
with Figure \ref{fig:fig4p1} and Figure \ref{fig:fig6p1}, we can see that the gap depths
 and particle concentration are qualitatively similar between 2-D and 3-D simulations. 

\begin{figure*}[ht]
\centering
\includegraphics[width=0.8\textwidth]{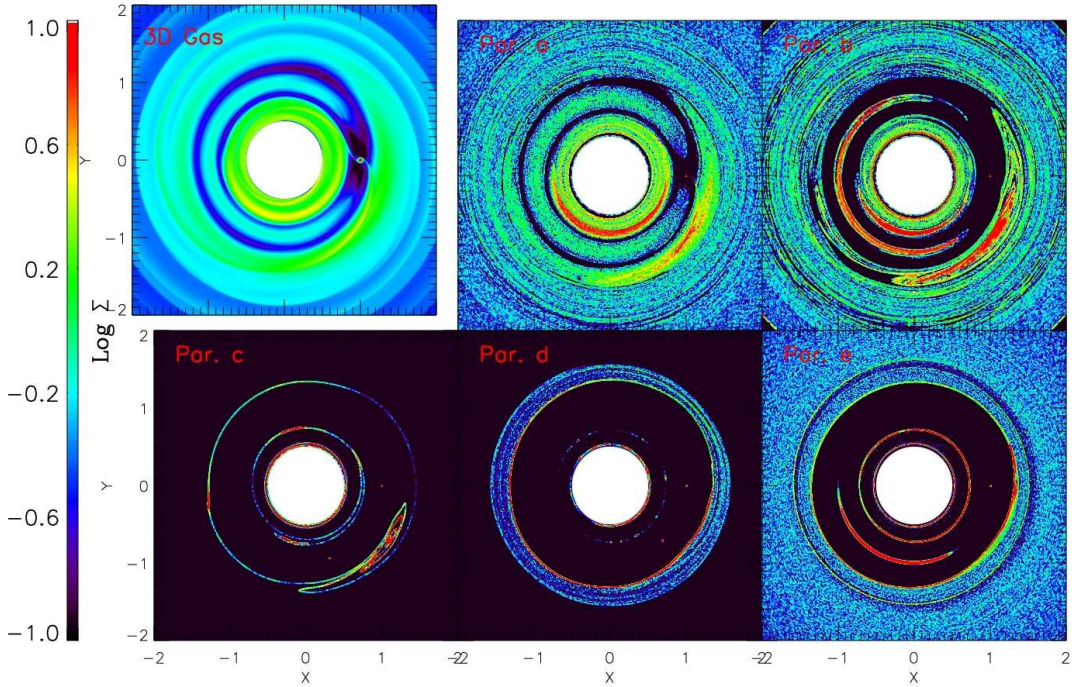} 
\vspace{-0.3 cm}
\caption{The gas and dust
surface density for the M50D3 case at 100 orbits in the X-Y plane. The vortex at the gap
edge is highly elongated along the $\phi$ direction.} \label{fig:fig11}
\end{figure*}

The gas and particle surface density for the M50D3 case in the x-y plane is shown
in Figure \ref{fig:fig11}. We can see that the vortex is 
highly elongated along the $\phi$ direction. 

\begin{figure*}[ht]
\centering
\includegraphics[width=0.8\textwidth]{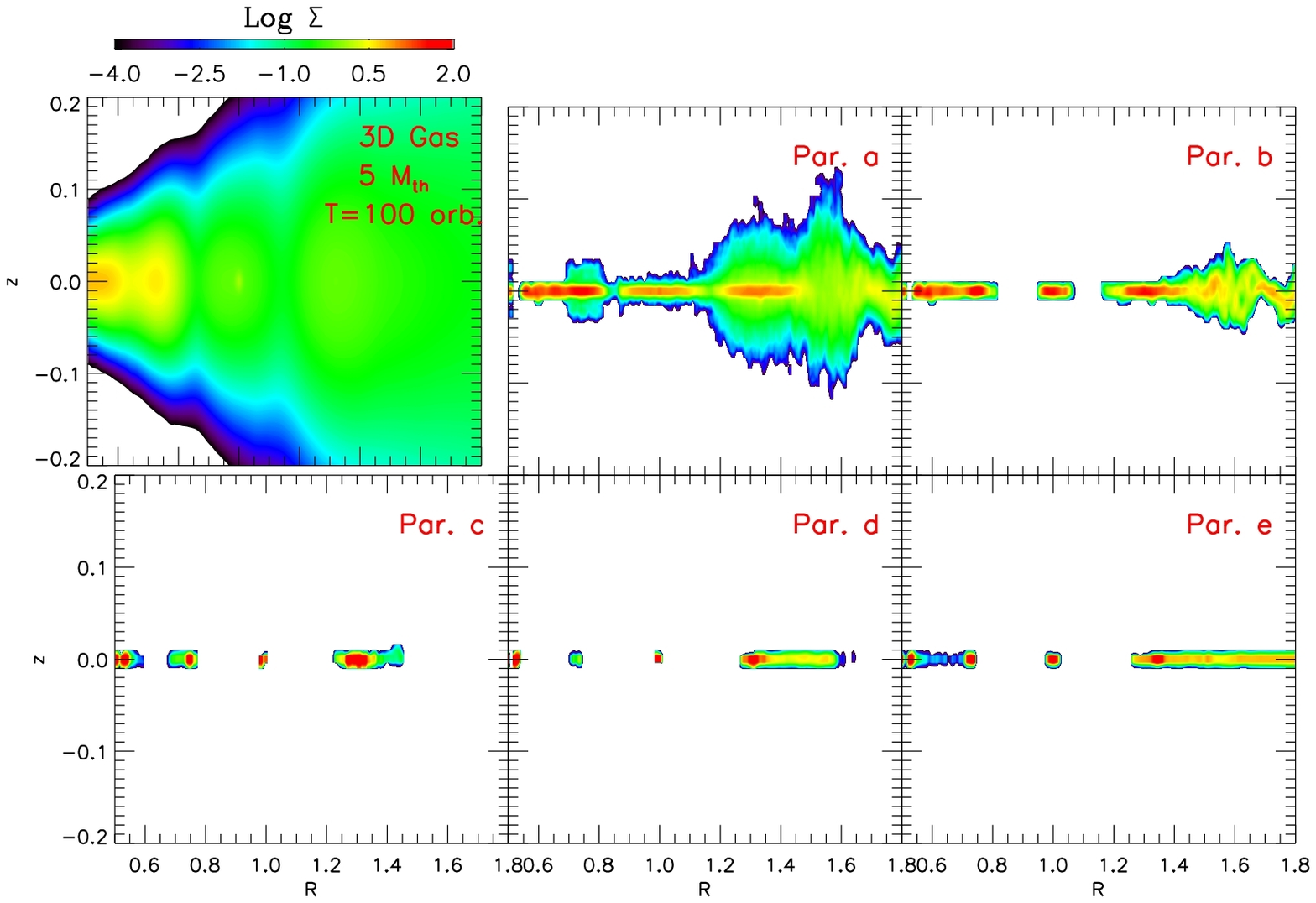} 
\vspace{-0.3 cm}
\caption{ The azimuthally averaged gas and dust 
 density for the M50D3 case at 100 orbits in the $R-z$ plane.  Particles have significant settling to 
 the midplane even in the vortices at the gap edges. 
} \label{fig:fig9p1}
\end{figure*}

\subsubsection{Particle Concentration in the Vortex}
Two outstanding questions regarding particle concentration in the vortex are
1) does the vortex have strong vertical motions which can stir up the particles, inhibiting
 settling and concentration
 within the vortex?
 2) particles of what size can be trapped in the vortex? If particles with quite different
 sizes can all be trapped within the vortex, the total dust mass inside the vortex can be
 high and planetesimal formation may be much easier. 
 
 \begin{figure}[ht]
 \centering
\includegraphics[width=0.5\textwidth]{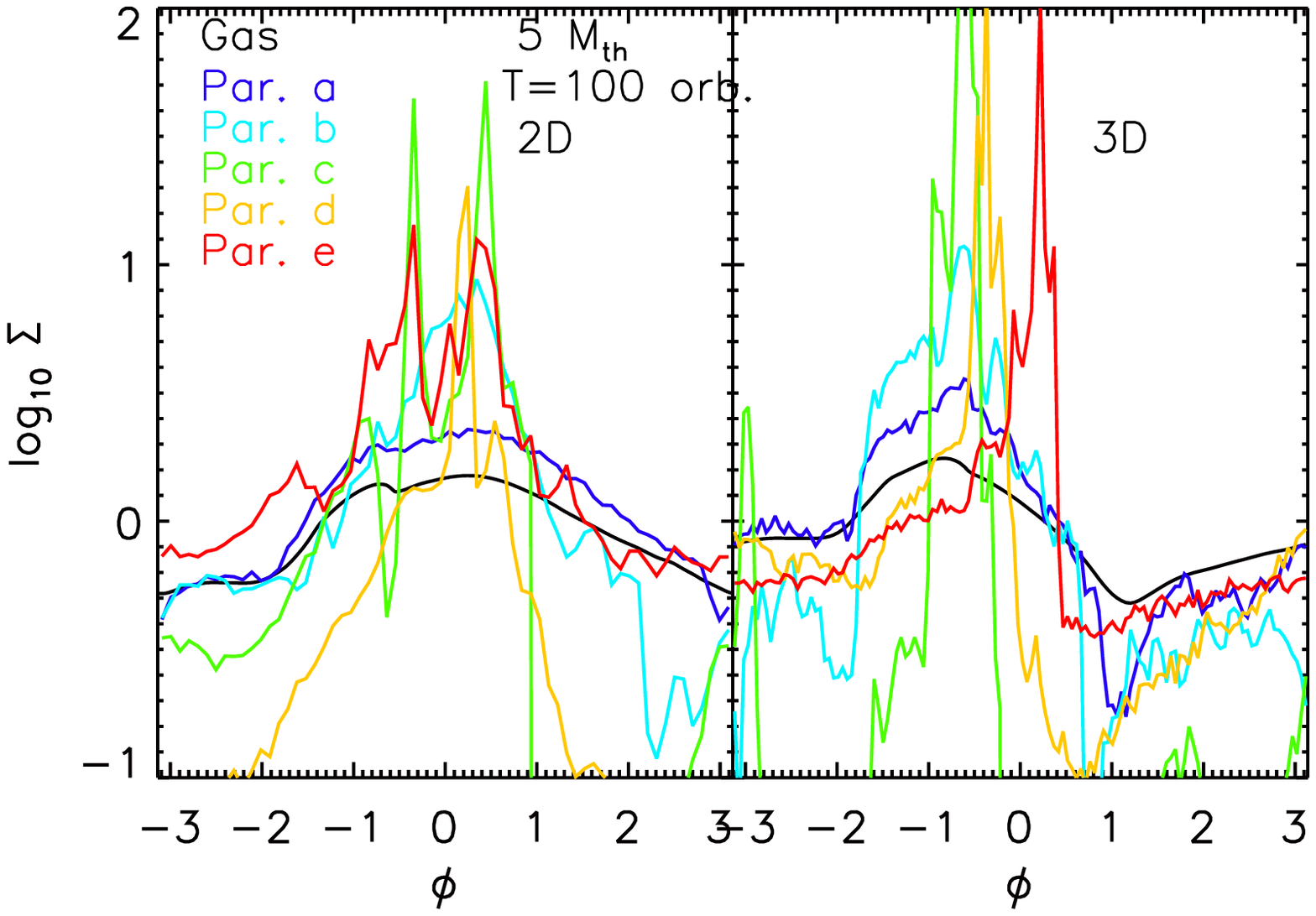} 
\vspace{-0.5 cm}
\caption{The gas (black curves) and particle (colored curves)
 surface density across the vortex center ($R\sim$1.32)
at the gap edge  along the $\phi$ direction for the 2-D (M50D2, the left panel) 
and 3-D simulation (M50D3, the right panel). The surface density shown is derived by averaging
the disk surface density from R=1.27 to R=1.37. The center of the vortex has the highest surface density.
Although the gas surface density only increases by a factor of 2-3 to the center of the vortex, it can concentrate dust
particles by a factor of 100. All particle species (except Par. a) have significant concentration within the vortex.
 } \label{fig:fig7p1}
\end{figure}
 
As shown in Section 4.1 and Figure \ref{fig:fig12}, the  
vortex in 3-D simulations does not have
noticeable  vertical motions within 2 scale heights.
Thus, without 
strong vertical motions at the midplane, the vortex cannot stir 
particles up to the atmosphere, and particles should settle towards the disk midplane
significantly. This expectation is confirmed in Figure \ref{fig:fig9p1}, where all types of particles
except Par. a have settled to the disk midplane within the vortex at the inner and outer gap edges
(The inner edge vortex is at $0.7<R<0.8$, and the outer edge vortex is at $1.2<R<1.4$
based on Figure \ref{fig:fig9}). Beyond the outer vortex at $R\sim 1.4$, there are some disturbances
at the midplane to stir up the particles. These disturbances seem to come from 
 the spiral density waves from the planet and waves excited by the vortex. The 
 simulation also suggests that the lower the density of a region is 
 (either at the surface or at the outer disk), the higher the velocity disturbance it has.
 In this sense,
 the center of the vortex where the density is highest and the disturbance is smallest
 acts as a quiescent  ``island'' allowing particles to settle. This result seems to be different from
  Meheut \etal (2012b) who claimed that big particles are stirred to the disk atmosphere by 
  the strong vertical motion in the vortex.

In a series of simulations by Meheut   \etal (2010, 2012a, 2012b), they reported strong vertical motions similar to ``convection rolls'' within the vortex. 
  To try to reproduce these results, we set up the disk structure with a density bump in the same way as Meheut \etal (2012a), using 
  the same resolution as used elsewhere in this paper. We find excellent agreements between our simulations
  and analytical calculations of both the RWI growth rate and eigenfunctions. Since we include real passive dust particles
  in our simulations, we can trace the trajectory of each individual particle. However, we found that dust particles 
  in the vortex only experience a small net vertical drag from the gas when they orbit around the vortex
  center. Such a small vertical drag will not stop the vertical settling of particles when the particle's $T_{s}$ is larger than $10^{-3}$
  (Zhu \& Stone, In prep). This is consistent with our current finding that all particles in the vortex (Table 1, $T_{s}>10^{-3}$) 
  settle to the disk midplane.

To quantitatively measure particle concentration 
within the vortex, we calculate the gas and particle
surface density along the $\phi$ direction  across the vortex center 
 in both 2-D and
3-D simulations (M50D2 and M50D3) at 100 orbits. 
Figure \ref{fig:fig7p1} shows such azimuthal density profiles at $R\sim 1.32$. The highest
surface density along the curve corresponds to the center of the vortex. While the gas surface
density increases by a factor of 2-3 towards the vortex center, the dust surface density
can go up by a factor of 10 for Par. b ($T_{s}\sim$0.02) and a factor of 
100 for Par. c ($T_{s}\sim$0.2). Even for Par. d and Par. e which have $T_{s}\sim$ 2 and 20, there is
significant dust concentration within the vortex. These big particles seem to concentrate at the region
which is slightly ahead of the vortex in the $\phi$ direction. 

In order to accumulate particles having $T_{s}\ll 1$ or $T_{s}\gg 1$, the vortex needs
to orbit around the central star at a speed which is very close to the background Keplerian 
speed. In this case, dust particles which also orbit at the Keplerian speed can stay 
in the vortex for many orbits and gradually
drift towards the vortex center azimuthally.  If the vortex orbits around the central star
at a speed which is different from the Keplerian speed, the particle trapped in with the vortex
will also move in a non-Keplerian speed. Thus, in the Keplerian reference plane, particles 
will experience a Coriolis force which has to be balance by the drag force induced by the vortex
rotation (Youdin 2010). When the deviation from the Keplerian speed is large enough, the vortex's 
rotation can
not provide enough drag force to balance the Coriolis force, 
and particles cannot be trapped and co-move with the vortex.

\begin{figure}[h]
\includegraphics[width=0.5\textwidth]{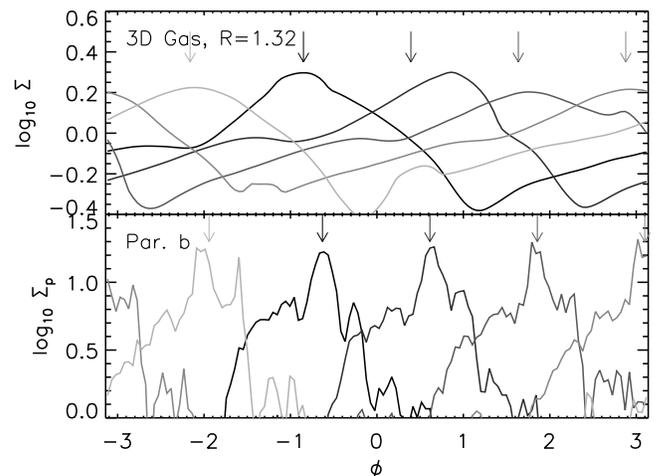} 
\vspace{-0.5 cm}
\caption{The gas (upper panel)
and Par. b (lower panel) surface density across the vortex center ($R=1.32$) at the gap 
edge along the $\phi$ direction at different times (from $t=628.3$ to $635.9$ 
with the interval of $\Delta t=$1.885) in the 3-D simulation (M50D3). The arrow labels
 where the vortex center would be if the vortex orbits at the background Keplerian speed at $R=1.32$.
} \label{fig:cut2devolve}
\end{figure}

\begin{figure}[h]
\includegraphics[width=0.5\textwidth]{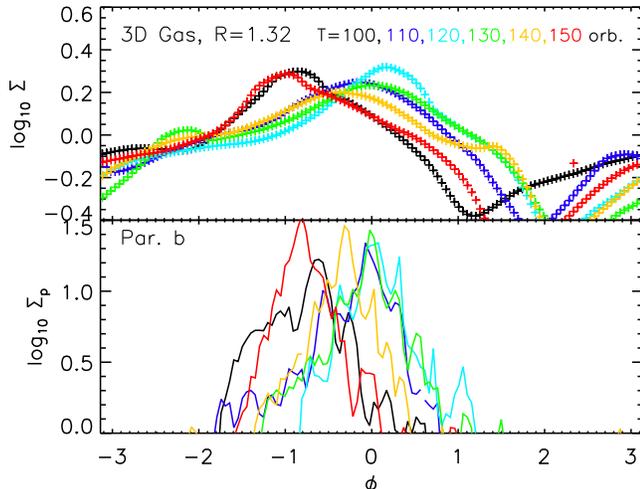} 
\caption{Similar to Figure \ref{fig:cut2devolve} but at a much longer timescale. The colored curves correspond
to the snapshots at $t=[100, 110 ... 150]\times 2\pi /\Omega$). Each curve is shifted
by the distance the vortex traveled if it orbits at the Keplerian speed at $R=1.32$. The closeness
of these curves suggests that the vortex orbits around the star at a highly Keplerian speed.
} \label{fig:cut2devolve2}
\end{figure}

Figure \ref{fig:cut2devolve} shows the disk surface density across the
vortex center at different times within one orbit. The arrows label the positions if
the vortex were to orbit around the central star at the Keplerian speed of its center. 
Although the highest
surface density in the gas may deviate from the Keplerian speed occasionally, there is no
 systematic drift of the vortex speed with respect to the Keplerian speed within one orbit. On a much
 longer timescale,  Fig \ref{fig:cut2devolve2}
 shows the disk surface density across the vortex center at T=100, 110, ...,150
  $\times 2\pi/\Omega(R=1)$ after shifting each curve by the distance the vortex travels if
  it were orbiting at a
  Keplerian speed at $R=1.32$. Again, the vortex
  is not perfectly steady but seems to follow the Keplerian orbit quite well. Assuming the distances  between the 
  peaks of these
  curves ($\sim 2/\Omega(R=1.32)$)
   are all due to the non-Keplerian rotation of the vortex within these 50
   orbits, we can estimate 
that the vortex's orbit deviates from the Keplerian orbit at most to a degree of
\begin{equation}
\left|\frac{v_{vortex}-v_{kep}}{v_{kep}}\right|<\frac{2/\Omega(R=1.32)}{50\times 2\pi/\Omega(R=1)}\sim 1\%\,.\label{eq:eqvortex}
\end{equation} 
Generally, throughout our simulations, the vortex follows the Keplerian orbit very well and particles with
a large range of sizes can be trapped in the vortex. Using the azimuthal drift velocity equation from Birnstiel
et al. (2013) and the gas velocity $v_{g,r}\sim 0.03 v_{kep}$ at the vortex edge from Figure \ref{fig:fig12},
we estimate that particles with $0.1<T_{s}<10$ can drift a distance R azimuthally within our simulation timescale of 100
orbits, which is consistent with Par. b to Par. e ($0.02<T_{s}<20$)  having some concentration in the vortex. 

 We want to note that it is possible the vortex orbits at a slightly non-Keplerian speed (below
$1\%$ deviation from the Keplerian speed) following the background gas flow.
It is known that,
due to the  radial pressure gradient, the background disk will either
rotate slightly sub-Keplerian (e.g.  the pressure decreasing with the radius), or slightly super-Keplerian (e.g. at the gap outer edge).
However, the background flow's deviation from the Keplerian speed is so small 
(e.g. with our disk parameters, the deviation from the Keplerian speed is only 0.125$\%$)
that we cannot tell if the vortex follows this slightly non-Keplerian background flow.
If the vortex has a slightly sub-Keplerian speed, it may be able to explain the particles' concentration 
ahead of the vortex in the azimuthal direction \footnote{Using Eq. 4.17 of Youdin (2010), $1\%$ deviation from the Keplerian speed
can lead to $\Delta \Phi$=0.1 with the particle's $T_{s}=0.1$ and the vortex's aspect ratio of 10.}.

\subsection{Weakly Coupled and Decoupled Particles}
\begin{figure*}[ht]
\centering
\includegraphics[width=0.8\textwidth]{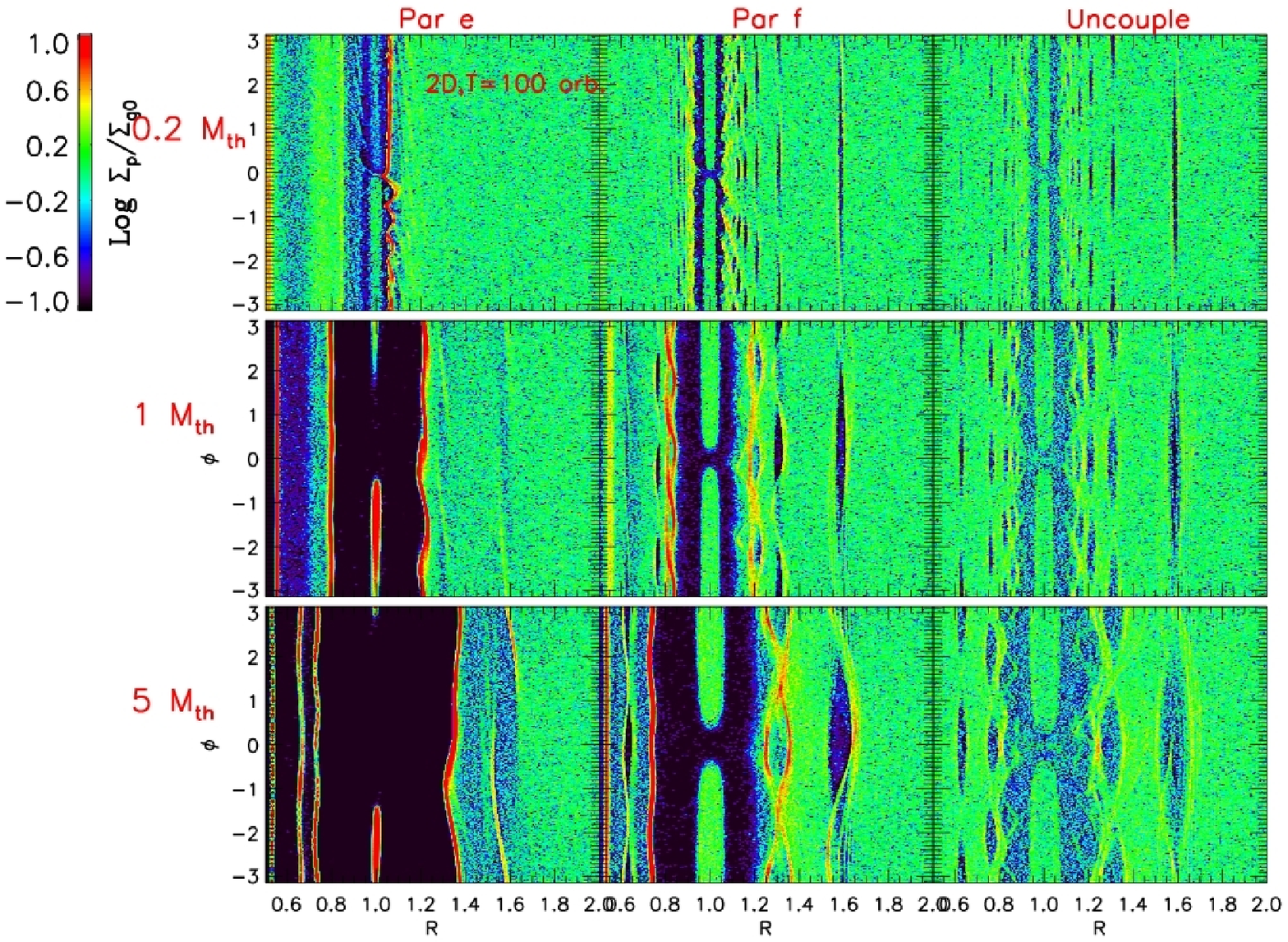} 
\vspace{-0.3 cm}
\caption{Particle surface density
for particle types which are in the weakly coupled and decoupled limit (Par. e to Par. g from left
to right) in 2-D simulations
with different planet masses (from top to bottom:
M02D2, M10D2, M50D2).   } \label{fig:fig13}

\centering
\includegraphics[width=0.8\textwidth]{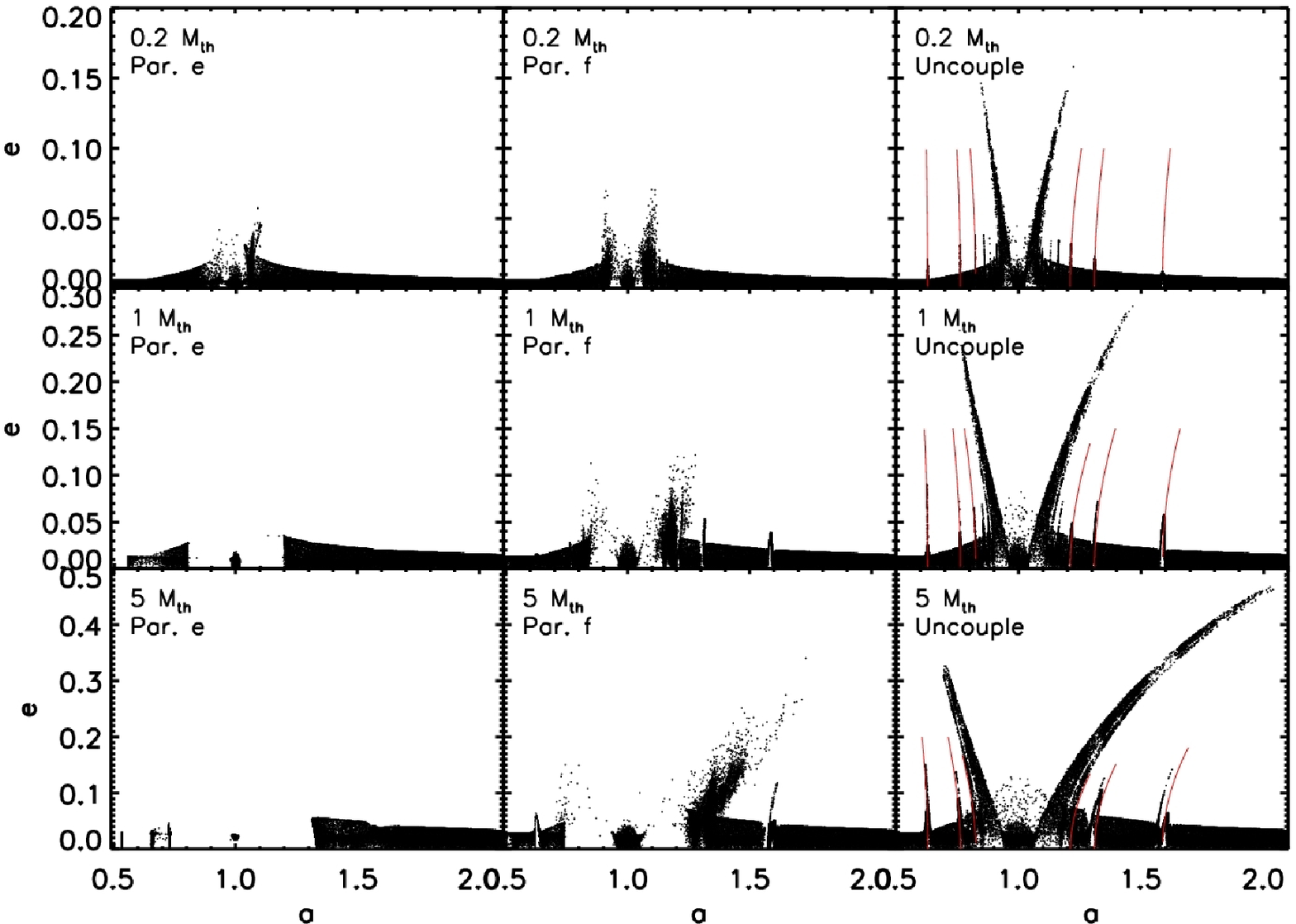} 
\vspace{-0.1 cm}
\caption{Similar to Figure \ref{fig:fig13}, but showing semi-major axis and eccentricity of particles.
 Gas drag can cause
particles radial drift and also damp particles eccentricities. 
Even if there is no significant gaps in the gas (e.g. M02D2),
a gap with 1-2 Hill radius wide can still be opened for weakly couple particles since gas drag damps the eccentricity of particles passing by the planet. 
If there is a gap in the gas (e.g. M50D2), particles are also trapped at the gap edges. 
The curves of constant Tisserand parameter are drawn in the rightmost panels as the red curves. } \label{fig:fig13p1}
\end{figure*}
\begin{figure*}
\centering
\includegraphics[width=0.8\textwidth]{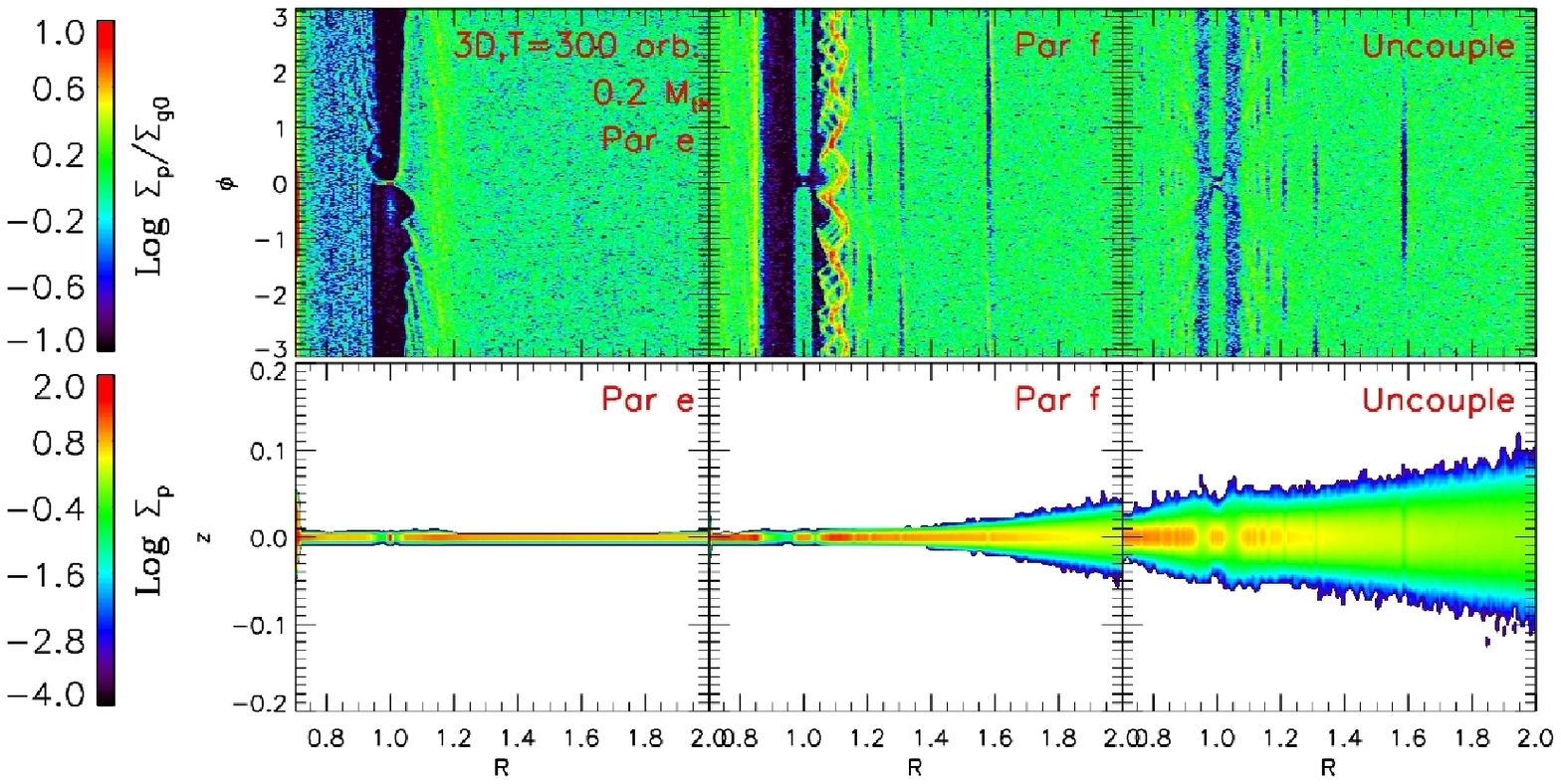}
\vspace{-3.2 cm}
\caption{Particle
 density for particle types which are in the decoupled limit in the  3-D simulation
 M02D3. Both disk density in the $R-\phi$ and $R-z$ planes
are shown. 
}\label{fig:figun0p2}

\centering
\includegraphics[width=0.8\textwidth]{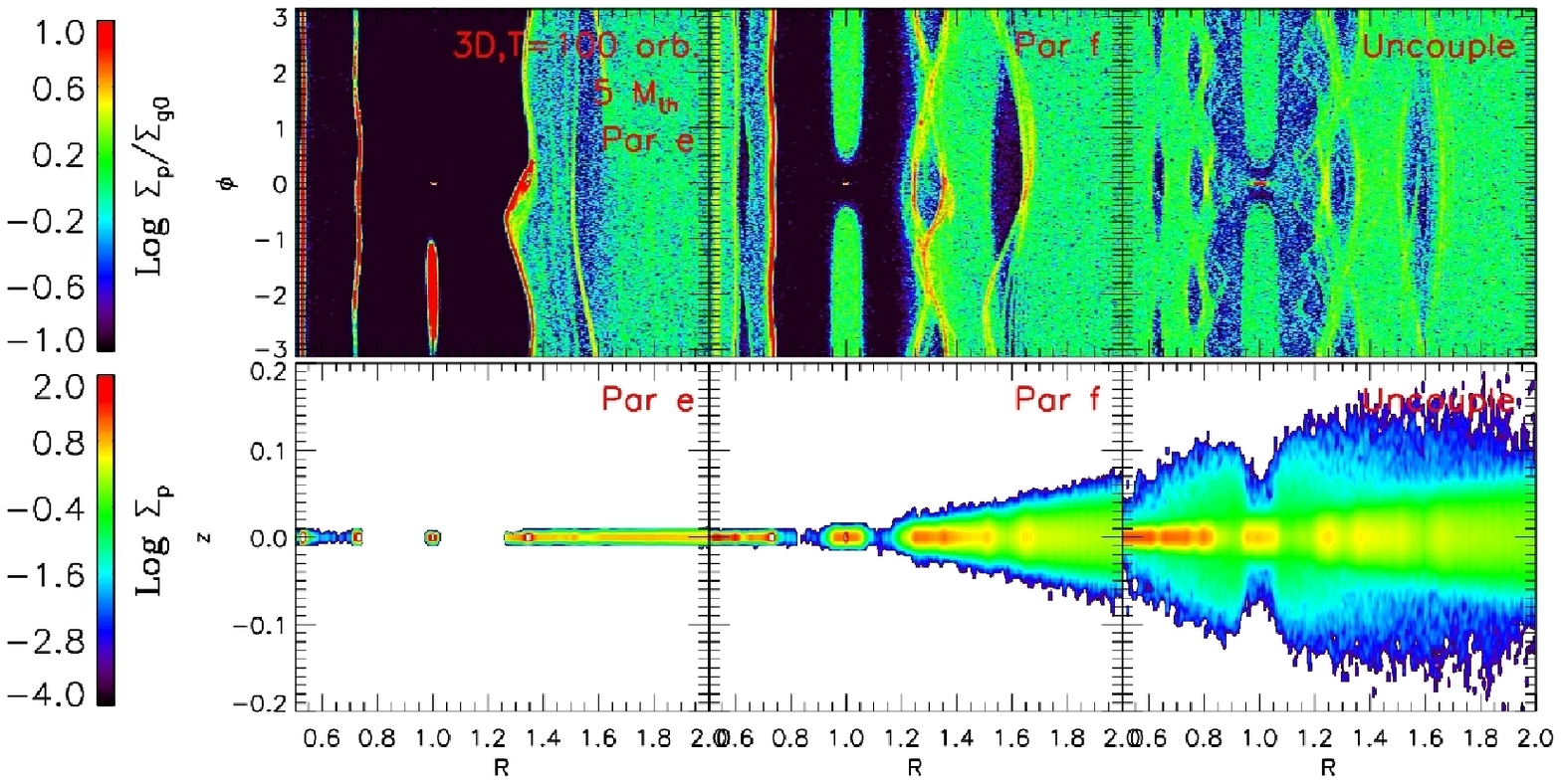} 
\vspace{-3.2 cm}
\caption{Similar to
Figure \ref{fig:figun0p2} but for the M50D3 case.
}\label{fig:figun5}
\end{figure*}
\subsubsection{2-D disks}
When the dust stopping time $T_{s}>1$, the eccentricity citation for particles
close to the planet and at mean motion resonances becomes visible.
Particles having the same eccentricity collectively
form ``ripples'' in the disk. Figure \ref{fig:fig13} shows the dust surface density for Par. e ($T_{s}\sim 20$), 
Par. f ($T_{s}\sim 200$), and Par. g (Uncoupled) in 2-D simulations M02D2 (top panels),
 M10D2 (middle panels), and M50D2 (bottom panels). Each ``ripple'' in these figures
 is produced by particles finishing one eccentric orbit. The bigger the ``ripple'', 
 the higher the eccentricity of the particles. 
 For Par. e, the eccentricity damping
 by gas drag is apparent in the M02D2 case. This decreasing amplitude of the ripple is
 similar to the ripple excited by Pan in Saturn's rings (Cuzzi 
\& Scargle 1985, Hedman et al. 2013) where the eccentricity damping is due to the particle collisions.
 Such eccentricity damping can reduce the particles' semi-major axis,
 eventually leading to a clean gap close to the planet or at mean-motion resonances, as shown in
 the Par. e and Par. f panels.
 In M10D2 and M50D2 cases, the sharp
  edge of the gap in the gas can still trap Par. e. This gap edge trapping becomes less 
 effective for less coupled particles (e.g. Par. f). 
 For fully decoupled particles (Par. g), eccentric orbits at the mean motion resonance 
become very apparent, which is also clearly shown in Ayliffe \etal (2012). 
At $R\sim 0.6$ and $R\sim 1.6$, the ripples correspond to 1:2 and 2:1 mean motion resonance, and 
at $R\sim 0.76$ and $R\sim 1.3$, they correspond to 3:2 and 2:3 mean motion resonance.
Generally, for first order resonances, 
$N$ ``ripples'' corresponds to the $(N+1):N$ resonance if the resonance
is outside the planetary orbit or the $(N-1):N$ resonance if the 
resonance is inside the planetary orbit.

Particle semi-major axis and eccentricity 
distributions for all the panels in Figure \ref{fig:fig13} are
shown in Figure \ref{fig:fig13p1}. Clearly, gas drag significantly damps
the eccentricity of particles close to the planet or at the mean motion
resonances.  It also facilitates particles gap opening around the planet.
This gap opening is the combination of dust trapping at the gap edge
and the shrinking of particle orbits due to the eccentricity damping. Similar
effect has been discussed in Tanaka \& Ida (1996,1999) and Rafikov (2001).

With little gas drag, eccentricity of big particles is continuously pumped up by the planet at mean-motion resonances.
In 2-D cases, particle eccentricity ($e$) and semi-major axis ($a$) follow the trajectories that keep the Tisserand parameter (Murray 
\& Dermott 1999) constant.
\begin{equation}
e=\left[1-\frac{a_{p}^{3}}{a}\left(\frac{1}{2a_{m}}-\frac{1}{2a}+\sqrt{\frac{a_{m}}{a_{p}^{3}}}\right)^{2}\right]^{1/2}\,,
\end{equation}
where $a_{p}$, $a_{m}$ are the semi-major axis of the planet and the mean-motion resonances. The curves
of constant Tisserand parameter at mean-motion resonances are shown in the rightmost panels of Figure \ref{fig:fig13} as the red curves.

\begin{figure*}[ht]
\centering
\includegraphics[width=0.8\textwidth]{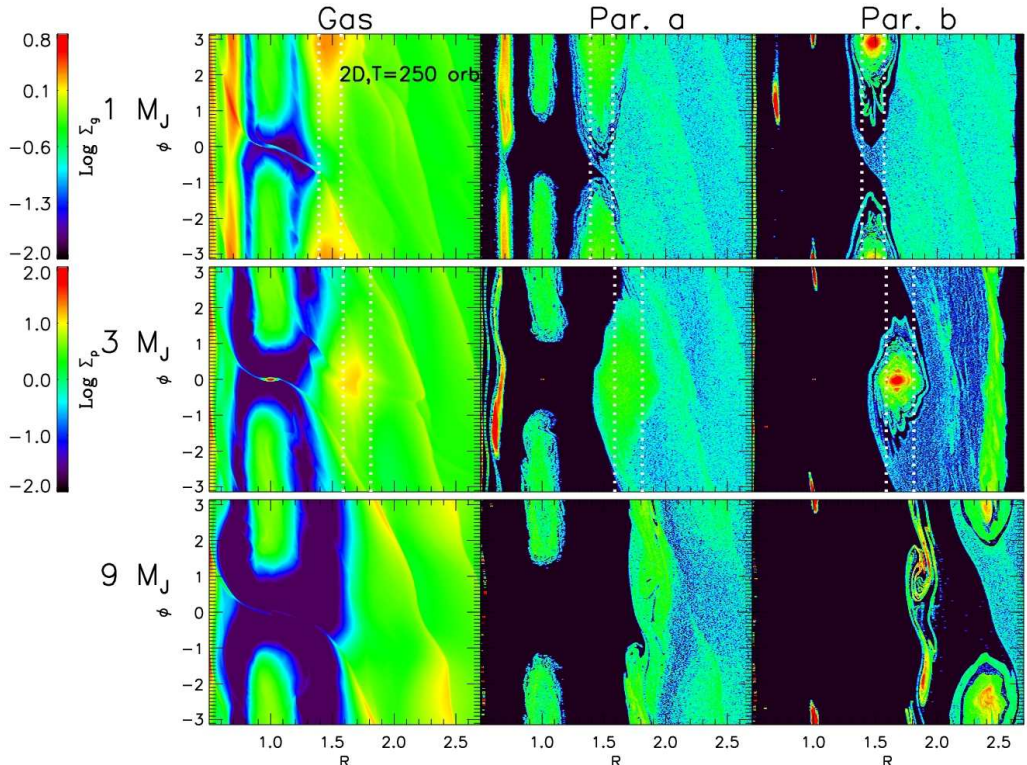} 
\vspace{-0.3 cm}
\caption{Disk surface density for gas (left), Par. a (middle), Par. b (right)
when there is a 1, 3, 9 M$_{J}$ mass planet (from top to bottom panels) at $R=1$ in the disk. The color bars
for gas and particle surface density are different for better image contrast. The dotted vertical lines in the 
upper two panels
label the disk radii which are one disk scale height away from the vortex center. Thus, the distance between
these two lines are 2 disk scale heights.} \label{fig:vamp}
\end{figure*}
\subsubsection{3-D disks}

The dust distributions in the $R-\phi$ plane and the $R-z$ plane for M02D3
and M50D3 cases are
shown in Figure \ref{fig:figun0p2} and Figure \ref{fig:figun5}. The dust distributions
in the $R-\phi$ plane are quite similar between 2-D and 3-D cases.
However, in 3-D disks, not only the particle eccentricity but also the
inclination can be excited if gas drag is small. In the lower right panels of Figures
\ref{fig:figun0p2} \& \ref{fig:figun5}, we can see that the disk is significantly
puffed up close to the planet. Gaps at the mean motion resonances
can also be seen in the $R-\phi$
panels.  However, any small amount of gas damping reduces the inclination of these
particles (Par. e and Par. f panels).

\section{Discussion}

\subsection{Massive Planets}
Considering the great similarity between 2-D and 3-D simulations, we use 2-D simulations
to explore how particles
are trapped in the gap edge vortex induced by a more massive planet.
Three simulations with a 1, 3, and 9 $M_{J}$ planet in the disk are carried out. Figure \ref{fig:vamp}
shows that a more massive planet opens a wider gap so that the gap edge vortex is further away
from the central star. The vertical dotted lines in the upper two panels show the positions in the disk which
are one disk scale height away from the vortex center. In other words, they label where the disk becomes supersonic with respect the the vortex center.
Clearly, the particle concentration can at most be 4 disk scale heights wide in the radial direction.

 With a 9 $M_{J}$ planet, the vortex at the gap edge at $R\sim 1.8$ is significantly weakened. However another vortex forms further out, at $R\sim 2.5$ away from the gap edge. By examining this new phenomenon in more detail, we find that its origin might be linked to disk evolution driven by the density waves launched at the low order ($m=2,3$) Lindblad resonances located far away from the planet. These resonances gain prominence for high-mass planets for two reasons. First, their strength (the amount of torque excited at them) increases quadratically with the planetary mass, while the torque accumulated from higher order resonances gets suppressed because of gap opening at lack of material at $\sim h$ away from the planet. Second, the non-linearity of the main wake excited closer in is higher for higher-mass planet, leading to its faster dissipation and smaller effect on the disk surface density far from the planet. Density waves excited at low-$m$ resonances have relatively long radial wavelength and dissipate via steepen into shocks far away from the planet. There waves deposit their angular momentum to the disk, modifying its surface density in such a way that a density maximum appears. This produces a vortensity minimum, which ultimately leads to RWI and vortex formation, as observed here. On the other hand, this secondary vortex far away from the gap edge also seems to be related to the primary vortex at the gap edge. It is also possible that the primary vortex at the gap edge excites density waves which steepening to shocks at large radii, leading to density pileup and vortex formation before the primary vortex disappears. Whether the secondary vortex
 is a transient effect and what is its detailed formation mechanism need further studies.
 
 Overall, a very massive planet in an inviscid disk is capable of inducing vortex whose distance 
from the central star is more
than twice the distance of the planet from the star. However the vortex is highly elongated along the $\phi$ direction in all these cases. 
Its radial extent is at most  0.2 $R$, while
azimuthally it can reach half of $2\pi$ radian.

\subsection{Comparison with Previous Work}
Paardekooper \& Mellema (2004, 2006) have carried out first 2-D two fluid simulations 
aimed at studying particle response to the presence of the planet. The fluid approach
for treating dust particles limited them to study
 grains with stopping time $T_{s}\sim 0.0005-0.02$ in disks with low mass planets which
 are not capable of opening gaps. They found that a 0.05 $M_{J}$ (or 0.4 $M_{th}$) planet
 can open gaps for particles
 having $T_{s}>0.005$ 
 after 100 orbits, and that some small particles concentrate at the 
 planet's mean-motion
 resonances. Although  our 2-D and 3-D simulations with 
 a 0.2 $M_{th}$ planet also open gaps for dust with T$_{s}\ge 10^{-2}$,
 we find that these dust gaps nicely correlate with the two gaps in the gas
opened by nonlinear wake steepening. No correlation between gaps for 
$T_{s}<1$ dust 
and planet's mean-motion resonances has been observed.
On the other hand, for particles with $T_{s}\gg 1$, we confirm that particles
get trapped in outer mean-motion resonances (Paardekooper 2007, Ayliffe \etal 2012). 

Focussing on a massive planet (5 $M_{J}$ or 50 $M_{th}$) in the disk, Fouchet \etal (2007, 2010)  
noticed dust concentration at the corotation region within the gap using SPH codes.
In our simulations, we point out that if the disk is inviscid, even a very low mass planet 
can cause the same effect.

On the other hand, all these simulations do not exhibit RWI at the gap edges.
This is either due to the shallowness of the gap opened by low mass planets (Paardekooper \& Mellema 2006), or
due to employing a relatively large
viscosity $\alpha=0.01-0.1$ (Paardekooper 2007; Fouchet \etal 2007, 2010; Ayliffe \etal 2012), 
which prevents the development of RWI. For viscous disks, dust diffusion due to
disk turbulence has been included in Zhu \etal (2012), which shows that dust diffusion can significantly
decrease the effect of dust trapping at gap edges. 
 
Lyra \etal (2009a) have carried out 2-D grid based hydrodynamic simulations with Lagrangian particles
to study planet's impact on dust particles in inviscid hydro disks, 
which is similar to our massive planet cases except that they run 2-D simulations with Cartesian 
grids. 
They found that a significant amount of particles are trapped in L4 and L5 Lagrange points and in the gap
edge vortices. With self-gravity
included, these particles can form earth mass planets. Our work extends Lyra \etal (2009a): 
we have carried out both 2-D and 3-D simulations with cylindrical grids (cylindrical grids
have higher effective resolutions than Cartesian grids for capturing RWI, as pointed out by de Val-Borro et 
al. 2007 and Lyra \etal 2009a), and explore particles with a large size range ($10^{-3}-10^{3}$ compared with $10^{-2}-1$).
We also focus on the 3-D structure and the Keplerian orbit of the vortex. They are crucial for understanding
how particles are distributed in the vortex and particles of what size can be trapped.  
 
Dust trapping by vortices which are produced by RWI of a dense ring
has been studied by Meheut (2012) using the two fluid approach.  Although they only follow the simulation for 3 orbits, 
significant dust concentration was observed. However, strong vertical motions within 1 scale height
in their vortices (also in Meheut 2010, 2012), and thus strong stirring for dust particles have not been
observed in our simulations.
The reasons for the discrepancy will be discussed in more detail in Zhu \& Stone (in prep), where
we found that, although particles undergo vertical oscillations when they orbit around the vortex center, the net height they gained
over many orbits is very small. 
At a much longer timescale than 3 orbits, particles with $T_{s}>10^{-3}$ are likely to settle to the vortex midplane.

Our simulations suggest that any mass planet can open gaps in both the gas and dust based 
on the nonlinear density wave theory  (Goodman \& Rafikov 2001, Rafikov 2002ab),
as long as the planet is not migrating and the disk is invisicid, 
which is in contrast with previous works on gap opening
in dust disks. If a low mass planet (e.g. $M_{\oplus}$) can indeed open gaps in the gas, 
it will stop the replenishment of particles coming
from the outside of the gap. Pebble accretion (Johansen 
\& Lacerda 2010, Ormel 
\& Klahr 2010, Lambrechts 
\& Johansen 2012) will be significantly reduced and
gas giant may form with a smaller core (Morbidelli \& Nesvorny 2012).
 
\subsection{Caveats and  Future Perspective} 
There are two main caveats in our present work: 1) dust feedback is neglected and 2) MRI turbulence in 
the disk is not considered.
 
Dust feedback plays an important role in several regions of a disk having a planet. First, 
dust which drifts quickest in the disk/vortex also settles quickest in the disk.
Due to the settling, dust to gas mass ratio can reach one in the midplane of the disk,
which will trigger Kelvin Helmholtz instability (Weidenschilling 1980, Johansen \etal 2006)  and streaming 
instability (Youdin \& Goodman 2006, Johansen \& Youdin 2007) at the midplane if
 the dust feedback to the gas is considered. These instabilities lead to disk turbulence which
 can slow down radial drift of dust particles (Bai \& Stone 2010b). Second, the dust feedback at the gap edge could slow down
 particles' trapping there (Ward 2009). Finally,
the dust feedback may also change
 the structure of the vortex itself. 
 
 Turbulence is essential for gas accretion onto the central star.
It also causes particle diffusion in the disk, which slows down particle radial
 drift and concentration (Youdin \& Lithwick 2007). However, 
 the turbulent structure of protoplanetary disks is not known very well. It is unclear if the disk
 is threaded by large scale magnetic fields, which will also change RWI 
(Tagger \& Pellat, 1999;  Yu \& Lai 2013). If turbulent ``dead zone'' exists in protoplanetary disks,
it can cause sharp density jumps at the ``dead zone'' edges which can also trigger 
RWI (Varni{\`e}re 
\& Tagger 2006; Lyra \etal 2009b; Lyra \& Mac Low 2012). The ``dead zone'' can also be very massive or even 
subject to gravitational instability (Zhu \etal 2011). Vortices at the planet-induced gap edges are
very different in massive disks compared with low mass disks (Lin \& Papaloizou 2010, Lin 2012b, Lovelace \& Hohfeld 2012)
due to self-gravity.
 
Regarding the vortices, we need to understand why they can exist over hundreds
of orbits in simulations without being destroyed by the elliptical instability (Kerswell 2002, Lesur \& Papaloizou 2009).
Thermodynamics and buoyancy effects can also play important roles in vortex dynamics (Lin 2013).
How dust particles are affected by eccentric disks is also important (Hsieh \& Gu 2012, Ataiee \etal 2013).
Ataiee \etal (2013) have shown that, unlike vortex, eccentric disks cannot trap particles.
 
\subsection{Observational Implications}
Our simulations suggest that even a low mass planet (e.g. several $M_{\oplus}$) 
can open dust gaps/cavities in an inviscid disk 
(e.g. in the ``dead zone'') as long as the
planet is not migrating. This
implies that dust gaps/cavities can be very common in protoplanetary disks, in agreement with the observation of
a large number of protoplanetary disks with submm cavities (Andrews \etal 2011). 
Furthermore, a low mass planet has an almost unnoticeable 
effect on the gas disk, which may explain missing cavities in near-IR polarized scattered light images (Dong \etal 2012)
and moderate disk accretion rates in some transitional disks (Espaillat \etal 2012).

At the sharp gap edge carved out by a massive planet, vortex can trap dust particles efficiently.
The surface density of a certain size dust particles can increase by over a factor 
of 100 within the vortex (Figure \ref{fig:fig7p1}). This significant asymmetric dust concentration has 
been observed with CARMA, SMA (Isella et al. 2013) and ALMA (van der Marel et al. 2013) and can be fitted quite well with particle
concentration in the vortex (Lyra \& Lin 2013). Although the observed asymmetric features
seem to have a larger radial extent than that of the vortex in our simulations (Figure \ref{fig:fig11}),
the observations only marginally resolved these asymmetric features radially and future higher resolution observations
are needed to constrain the shape of these features. 
If the asymmetric dust feature is indeed caused by the vortex, it may imply 
the existence of a low-turbulent region (e.g. the ``dead zone'') in protoplanetary disks,
since the vortex formation is significantly suppressed in viscous disks (de Val-Borro et 
al. 2007, Zhu \etal 2012). However, MRI turbulent disks may behave quite differently 
from viscous disks regarding the vortex formation (Lyra 
\& Mac Low 2012). 
We are carrying MHD simulations to study the effect of MRI turbulence on the vortex formation.

Our simulations suggest that the vortex orbits around the central star at the background Keplerian
speed. Thus, if observed asymmetric features (van der Marel et al. 2013) are indeed caused by the gap edge vortex, we should be
able to detect the Keplerian motion in future (in tens of years). 

\section{Conclusion}
In this paper we have implemented Lagrangian particles into Athena using cylindrical
grids to simulate gas and dust dynamics simultaneously. 
We have performed a systematic study on how planets affect dust distributions in protoplanetary disks 
by carrying out both 2-D and 3-D inviscid hydrodynamic global simulations with dust particles in the
presence of a 
planet. The planet mass spans from well below the thermal mass ( 0.2 $M_{th}$, equivalent to 8 $M_{\oplus}$ )
 to well above the thermal mass (5 $M_{th}$, or 0.65 $M_{J}$ in 3-D simulations, and 9 $M_{J}$ in 2-D simulations).
The dust size
spans more than 6 orders of magnitude from the well-coupled to decoupled limits ($T_{s}\sim 10^{-3} - 10^{3}$).

In the presence of a low mass planet with $M_{p}<M_{t}$, two shallow  gaps in the gas
start to open on each side of the planet
at the distance where nonlinear spiral wakes steepen to shocks. This gap opening process
is robust in both 2-D and 3-D simulations. We find that although these gaps may be barely 
visible in the gas disk after several hundred orbits the gaps are very prominent in the dust
since even the shallow gaps in the gas change the drift speed of dust particles leading to particle
pile up at the gap edge. Particles also concentrate at the co-orbital region of the planet.

A more massive planet can induce gaps in the gas quicker. The sharp gap edges
are Rossby wave unstable and develop vortices which eventually merge
into a single vortex. With the continuous driving by the shock, the vortex 
can last more than several hundred orbits in our 2-D and 3-D simulations until the end of the simulations.
The vortex can effectively trap dust particles. We find that the vortex is intrinsically
2-dimensional having no strong vertical motion within 1 disk scale height, and thus cannot stir particles from the midplane
to high altitudes. Particles in the vortex settle to the midplane, similar to the rest of the disk.
We also find that the vortex orbits around the central star at nearly the background Keplerian speed. The deviation from
the Keplerian speed is less than $1\%$, which is crucial for secularly trapping 
particles with $T_{s}\ll 1$ and $T_{s}\gg 1$. In our simulations, particles with $0.02<T_{s}<20$ have
significant concentration within the vortex, although smaller particles trace the 
structure of the vortex much better than bigger particles. 

We explore how particles
are trapped in the gap edge vortex induced by more massive planets. 
A more massive planet opens a wider gap so that the gap edge vortex is further away
from the central star. A 9 M$_{J}$ planet induces a vortex whose distance from the central star is more
than twice the distance of the planet from the star.

For particles with $T_{s}\gg 1$, particle eccentricity is driven up by the planet and is gradually damped by the gas drag after
many orbits, forming ``ripples'' at the gap edge, similar to Pan in Saturn's rings. 
We also find significant eccentricity pumping at the mean motion resonances,
which facilitates gap opening. 

Planet gap edges and vortices can efficiently trap particles with a large range of sizes. Thus,
a significant amount of dust can be trapped potentially accelerating planetesimal and
planet formation. They should also provide distinctive features which can be probed with ALMA
and EVLA.

The limitations of this work include the lack of the turbulent structure
of the disk, particle feedback to the gas, and self-gravity of the dust particles. These effects will 
be studied step by step in our future work on this subject.

\acknowledgments
This work was supported by NSF grant AST-0908269 and Princeton University.
It is also supported in part by the National Aeronautics and Space
Administration under Grant/Contract/Agreement No. NNX10AH37G issued through the Origins of Solar Systems program.
All simulations were carried out using
computers supported by the Princeton Institute of Computational Science and Engineering and Kraken at National 
Institute for Computational Sciences through XSEDE grant TG-AST130002. 
Z.Z. thank Catherine Espaillat, Hanno Rein, Min-Kai Lin, and Subo Dong for helpful discussions.

\appendix
\section{A. Particle Integrators}

Three particle integrators have been developed to solve the equation of motion Eq. (\ref{eq:motion}) using cylindrical coordinates.
The drag term is added as in Bai \& Stone (2010a). Here we focus on the equation without the drag term 
\begin{equation}
\frac{d\mathbf{v}_{i}}{dt}=-\nabla\Phi\,.\label{eq:motion2}
\end{equation}
and designing integrators which conserve
the geometric properties of particle orbit under the gravitational force only.

$Leapfrog$ $in$ $Cartesian$ $Coordinates$ (Car SI): This integrator is implemented into Athena by Bai \& Stone (2010a).
In order to apply this integrator for particles in cylindrical grids, we first transform the particle position from cylindrical coordinates 
to Cartesian coordinates. Then we apply this integrator to advance  the particle position and velocity in Cartesian coordinates, 
and finally we transform these quantities back into cylindrical coordinates.

Due to leapfrog's simplicity, symplectic
and time-reversible properties, it is widely used in orbit integration (refer to Chap. 3 of Binney \& Tremaine 2008
for a review on symplectic integrators). 
However, we want to
emphasize that leapfrog is guaranteed to be symplectic  only in Cartesian coordinates with
a static potential where its Hamiltonian can be written as 
${\cal H}=(1/2)m( v_{x}^{2}+v_{y}^{2}+v_{z}^{2})-\Phi(x,y,z)$, which can be 
separated as ${\cal H}={\cal H}_{drift} (p)+{\cal H}_{kick}(q)$ 
(where $p$
and $q$ are canonical coordinates of one particle's motion in the phase space: $p=[mv_{x}, mv_{y}, mv_{z}] $
and $q=[x,y,z]$). The equation of motion derived from ${\cal H}_{drift} (p)$
is the drift step, and that derived from ${\cal H}_{kick} (q)$ is the kick step.
Leapfrog is also time-reversible since it can be rearranged as a half drift---full kick---half drift scheme.
Due to its
separable Hamiltonian, leapfrog can be extended to higher order symplectic schemes 
(Yoshida 1993, Forest 2006 ). Or we can 
directly transform the particle position and velocity from cylindrical coordinates to
orbital elements and use mixed variable integrators such as the Wisdom-Holman integrator (Wisdom 
\& Holman 1992).
These expensive higher order integrators are not needed for this work, but can be easily implemented
in future if desired.

However, the coordinate transformation uses $sin$ and $cos$ functions which are 
computationally expensive. A bigger issue is that the particle follows a zigzag path along the orbit during drift/kick steps.
The drag term is calculated at the kick step ($t=t+\Delta t/2$) when  the particle is farthest away from
its correct path. 
A better scheme is applying Velocity-Verlet integrators twice so that
we have correct velocity and position at time $t+\Delta t/2$ to calculate the drag terms. However, such a method will double the
time for the particle integration.

$Leapfrog$ $in$ $Cylindrical$ $Coordinates$ (Cyl SI): Due to above limitations, we want to design an algorithm 
working directly under cylindrical coordinates. The Hamiltonian under cylindrical coordinates is
${\cal H}=(1/2)m (  v_{R}^{2}+ l^{2}/R+ v_{z}^{2})-\Phi(R,\phi,z)$, where $l=Rv_{\phi}$. The equations of motion are 
(for simplicity, we ignore $v_{z}$ in the following derivation.)
\begin{eqnarray}
\frac{d R}{d t}=v_{R} \,\,\,,\,\,\,\frac{d \phi}{d t}=\frac{l}{R^{2}}\\
\frac{d v_{R}}{dt}=\frac{l^{2}}{R^{3}}-\frac{\partial \Phi}{\partial r}\,\,\,,\,\,\,\frac{d l}{dt}=-\frac{\partial \Phi}{\partial \phi}\,.
\end{eqnarray}
Unfortunately ${\cal H}$ is not separable
for canonical coordinates ($R, \phi, z$ and $v_{R}, l, v_{z}$), and implicit methods are needed to 
design a symplectic implicit
integrator from ${\cal H}$ (Yoshida 1993) which will slow down the orbit integration significantly. 
Here, we still follow the spirit of leapfrog and separate the equation of motion
as drift and kick steps. In the following, we will show that, with a proper design, the scheme conserves phase-space volume so that it satisfies Liouville theorem. 
Furthermore, we can even make this scheme time-reversible by ordering the equations.
The variables are updated as follows:
\begin{eqnarray}
Half\,\, Drift:\,\,\,\,\,\,\,\,\,\,\,\,\,\,\,\,\,\,\,\,\,\,\,\,\,\,\,\,\,\,\,\,\,\,\,\,\,\,\,\,\,\,\,\,\,\,\,\,\,\,\,\,\,\,\,\,\,\,\,\,\,\,\,\,v_{R,n+1}=v_{R,n}\\
l_{n+1}=l_{n}\\
R_{n+1}=R_{n}+v_{R,n}\frac{\Delta t}{2} \\
\phi_{n+1}=\phi_{n}+\frac{1}{2}\left(\frac{l_{n}}{R_{n}^{2}}+\frac{l_{n+1}}{R_{n+1}^{2}}\right)\frac{\Delta t}{2}\\
Kick:\,\,\,\,\,\,\,\,\,\,\,\,\,\,\,\,\,\,\,\,\,\,\,\,\,\,\,\,\,\,\,\,\,\,\,\,\,\,\,\,\,\,\,\,\,\,\,\,\,\,\,\,\,\,\,\,\,\,\,\,\,\,\,\,\,\,\,\,\,\,\,\,\,\,\,\,R_{n+2}=R_{n+1}\\
\phi_{n+2}=\phi_{n+1}\\
l_{n+2}=l_{n+1}-(\frac{\partial \Phi}{\partial \phi})_{n+1}\Delta t \label{eq:eqlnp2} \\
v_{R,n+2}=v_{R,n+1}+\left[\frac{1}{2}\left(\frac{l_{n+1}^{2}}{R_{n+1}^{3}}+\frac{l_{n+2}^{2}}{R_{n+2}^{3}}\right)-(\frac{\partial \Phi}{\partial R})_{n+1}\right]\Delta t 
\label{eq:eqvrnp2} \\
Half\,\, Drift:\,\,\,\,\,\,\,\,\,\,\,\,\,\,\,\,\,\,\,\,\,\,\,\,\,\,\,\,\,\,\,\,\,\,\,\,\,\,\,\,\,\,\,\,\,\,\,\,\,\,\,\,\,\,\,\,\,\,\,\,\,\,\,\, The\,\,
same\,\,as\,\,the\,\, previous\,\, half\,\, drift
\end{eqnarray}
Please note that in the drift step $\phi$ has to be updated later than $R$, and in the kick step $v_{R}$ has to be updated
later than $l$ in order to make the scheme time-reversible. The above scheme conserves the phase-space volume, since the Jacobian matrix for variables ($R, \phi, v_{R}, l$) 
in the drift and kick steps are
$J_{drift} =\left(
\begin{array}{cccc}
1 & 0 & \Delta t /2& 0 \\
A & 1 & B & C \\
0 & 0 & 1 & 0 \\
0 & 0 & 0 & 1\\
\end{array} \right)$
and
$J_{kick} =\left(
\begin{array}{cccc}
1 & 0 & 0 & 0 \\
0 & 1 & 0 & 0 \\
D & E & 1 & F \\
G & H & 0 & 1\\
\end{array}\right)$, where $A$ to $H$ are non-zero quantities. It is obvious that the determinants for both of these
matrices are 1. It's easy to show that the Jacobian is still one with the z-component  position and velocity added in the leapfrog fashion. 
The Jacobian for the leapfrog in Cartesian coordinates have matrix components $A=0$ and $F=0$. Here we take advantage
of the special form of the Jacobian to make the scheme time reversible (adding more terms to A and F will not change $|J|=1$).  
Here we rely on the special form of this Jacobian matrix for
 designing a phase-volume conserving integrator. We cannot find a simple leapfrog scheme which has this property
either for the shearing box approximation (Rein \& Tremaine 2011) or for spherical coordinates. 

\begin{figure}[h]
\centering
\includegraphics[width=0.35\textwidth]{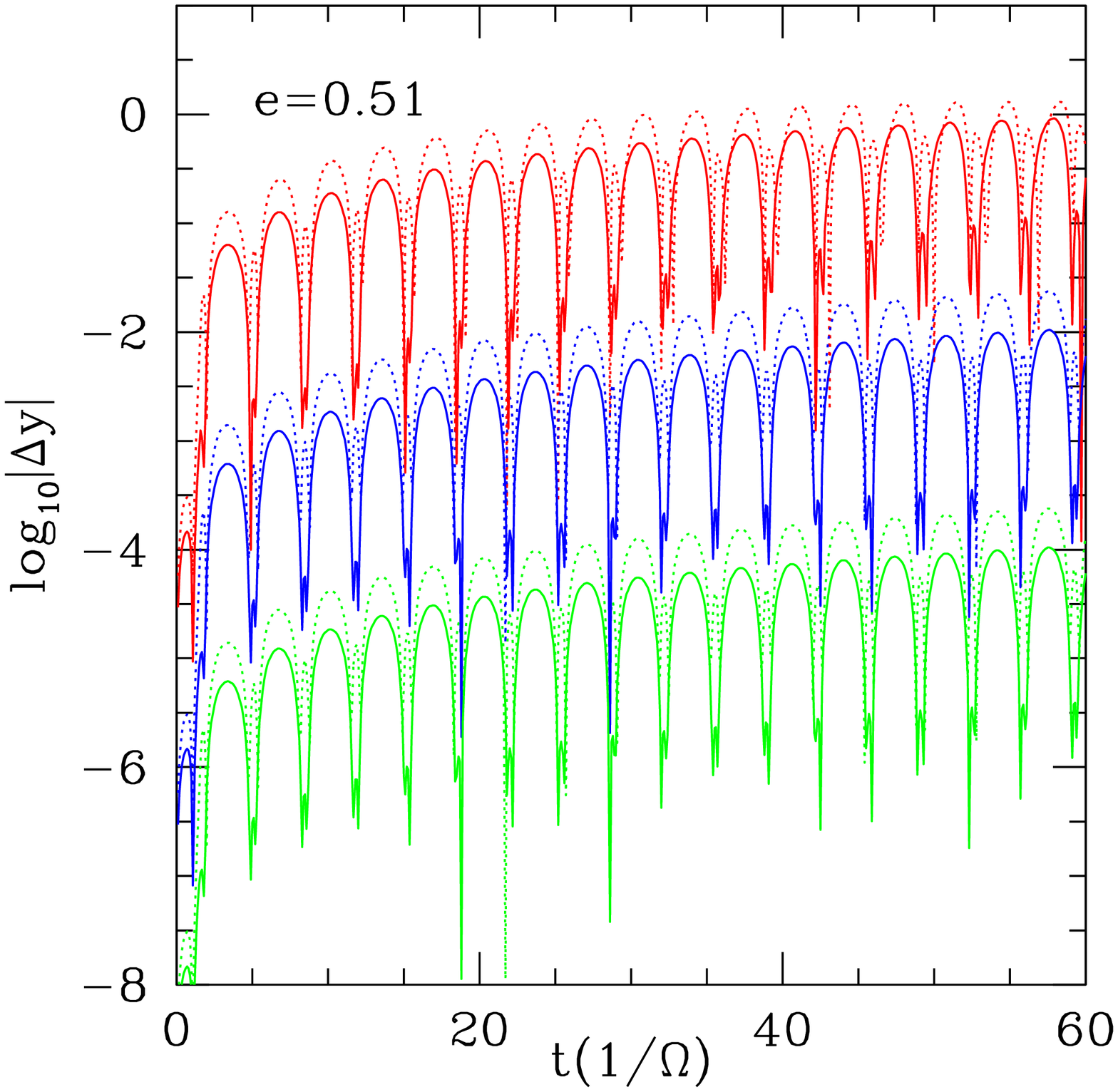} \
\includegraphics[width=0.35\textwidth]{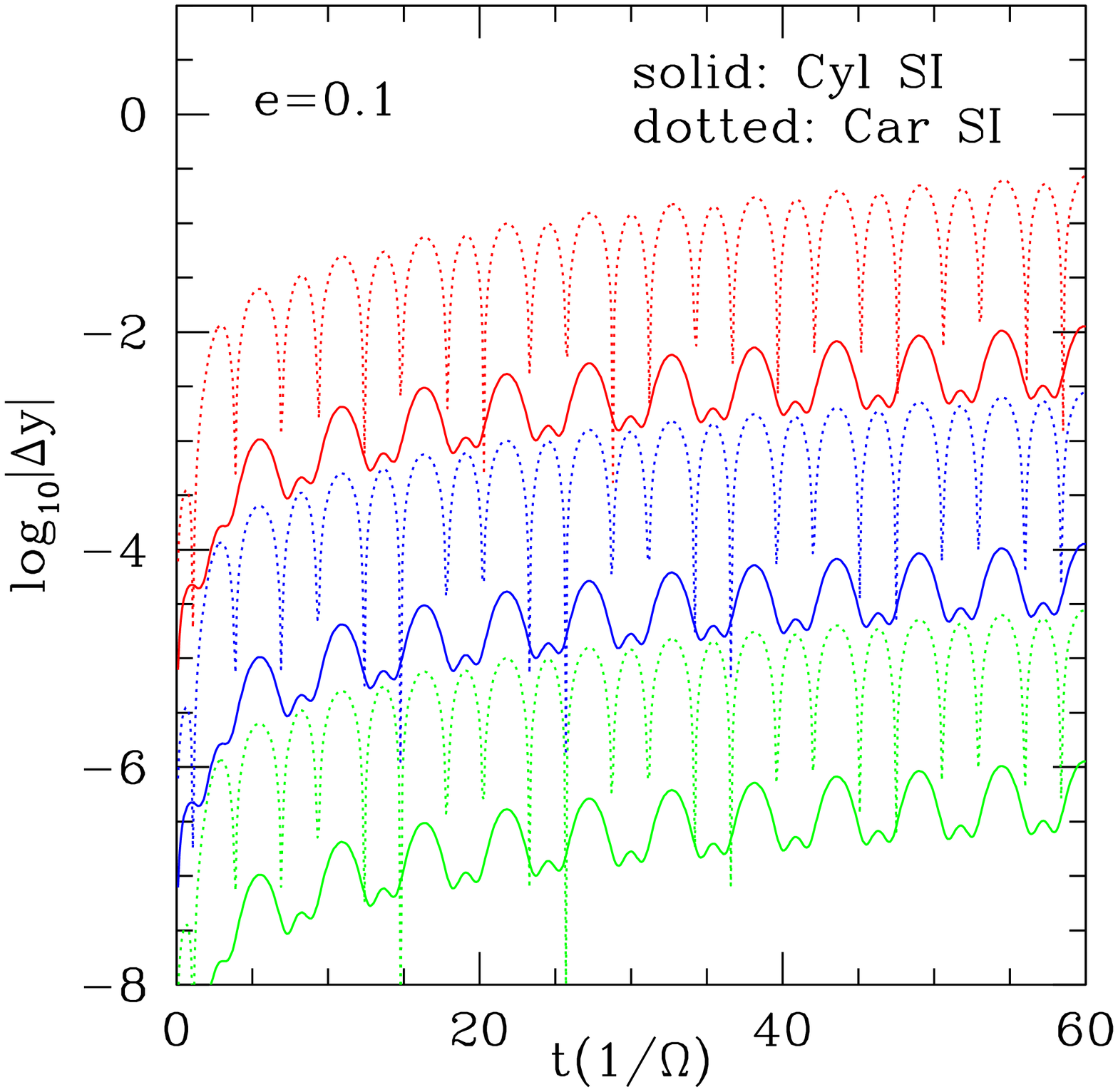} \\
\vspace{-0.1 cm}
\caption{The position error for integrating an  $e=0.51$ (the left panel)
and $e=0.1$ (the right panel) orbit using integrators Cyl SI (cylindrical leapfrog) 
and Car SI (Cartesian leapfrog) with a
time step of $\Delta t$=0.1 (red curves), 0.01(blue curves), 0.001(green curves).
These integrators are second order accurate. Cyl SI is significantly better
than Car SI for integrating particle orbit with a low eccentricity.
 } \label{fig:figorbits}
\end{figure}

This particle integrator is also second-order accurate and as simple as leapfrog in Cartesian coordinates
but without expensive coordinate transformation. It performs even better than Cartesian Leapfrog
for moderately eccentric orbits. Most importantly, the solution is exact if $e=0$
since the particle just drifts along the $\phi=\Omega\times t$ curve.
This is in contrast with leapfrog in Cartesian coordinates where the solution is exact
for a free floating particle ($e\to\infty$). Thus we expect that, for particle orbits having low eccentricity, 
our leapfrog
in cylindrical coordinates behaves far better than the traditional leapfrog in Cartesian
coordinates. As shown in Figure \ref{fig:figorbits}, this integrator has two orders 
of magnitude smaller position error than the leapfrog in Cartesian coordinates for 
integrating an $e=0.1$ orbit. Thus gas-dust coupling term is calculated
at a more accurate particle position compared with the most inaccurate particle position in the leapfrog in Cartesian coordinates, as long as
the eccentricity of the particle orbit is low (which is the case when dust and gas are moderately coupled).  
This property allows us to use almost one order of magnitude larger time step than the leapfrog in Cartesian coordinates to get the same
accurate particle drift speed (Appendix B). 

When the gas drag is included at the kick step (Bai \& Stone 2010a), $v_{r,n+2}$ (Eq. \ref{eq:eqlnp2}) and
$l_{n+2}$ for the kick step (Eq. \ref{eq:eqvrnp2}) become
\begin{eqnarray}
l_{n+2}=l_{n+1}+\frac{-(\frac{\partial \Phi}{\partial \phi})_{n+1}\Delta t+(v_{g,\phi,n+1}R_{n+1}-l_{n+1})\Delta t/t_{s,n+1}}{1+\Delta t/(2t_{s,n+1})}\\
v_{R,n+2}=v_{R,n+1}+\frac{\left[\frac{1}{2}\left(\frac{l_{n+1}^{2}}{R_{n+1}^{3}}+\frac{l_{n+2}^{2}}{R_{n+2}^{3}}\right)-(\frac{\partial \Phi}{\partial R})_{n+1}\right]\Delta t+(v_{g,R,n+1}-v_{R,n+1})\Delta t/t_{s,n+1}}{1+\Delta t/ (2 t_{s,n+1})}
\end{eqnarray}

$Fully$ $Implicit$ $Integrator$ $in$ $Cylindrical$ $Coordinate$ (Cyl IM): for particles with the stopping time significantly
shorter than the numerical time step, the drag term can dominate the gravitational force term and makes the integrator
numerically unstable. Thus, we develope a fully implicit integrator following Bai \& Stone (2010a). Unfortunately, such an integrator is
not phase-volume conserving. We will illustrate how we develop this integrator using $l$ as an example. First, as a fully-implicit
 integrator, we want to 
express $l_{n}$ using quantities at the $n+1$ step.
\begin{equation}
l_{n}=l_{n+1}+\left(\frac{\partial \Phi}{\partial \phi}\right)_{n+1}\Delta t+\frac{R_{n+1}}{t_{s,n+1}}\left(\frac{l_{n+1}}{R_{n+1}}-v_{g,\phi,n+1}\right)\Delta t\,.
\end{equation}
Then we plug $l_{n}$ to the right side of 
\begin{equation}
\frac{l_{n+1}-l_{n}}{\Delta t}=-\frac{1}{2}\left(\frac{\partial \Phi}{\partial \phi}\right)_{n}-\frac{1}{2}\left(\frac{\partial \Phi}{\partial \phi}\right)_{n+1}-\frac{1}{2t_{s,n+1}}(l_{n+1}-R_{n+1} v_{g,\phi,n+1})-\frac{1}{2t_{s,n}}(l_{n}-R_{n}v_{g,\phi,n})\,,
\end{equation}
to derive
\begin{equation}
l_{n+1}=l_{n}+\frac{-\frac{\Delta t}{2}\left(\frac{\partial \Phi}{\partial \phi}\right)_{n}-\frac{\Delta t}{2 t_{s,n}}(l_{n}-R_{n}v_{g,\phi,n})+(-\frac{\Delta t}{2}\left(\frac{\partial \Phi}{\partial \phi}\right)_{n+1}-\frac{\Delta t}{2t_{s,n+1}}(l_{n}-R_{n+1}v_{g,\phi,n+1}))(1+\frac{\Delta t}{t_{s,n}})}{1+\Delta t(1/ (2t_{s,n})+1/(2t_{s,n+1})+\Delta t/(2t_{s,n}t_{s,n+1}))}
\end{equation}
Similarly $v_{R,n+1}$ can be derived as
\begin{equation}
v_{R,n+1}=v_{R,n}+\frac{-\frac{\Delta t}{2}\left(\frac{\partial \Phi}{\partial R}\right)_{n}-\frac{\Delta t}{2 t_{s,n}}(v_{R,n}-v_{g,R,n})+\frac{\Delta t l_{n}^{2}}{2 R_{n}^{3}}+(-\frac{\Delta t}{2}\left(\frac{\partial \Phi}{\partial R}\right)_{n+1}-\frac{\Delta t}{2t_{s,n+1}}(v_{R,n}-v_{g,R,n+1})+\frac{\Delta t l_{n+1}^{2}}{2 R_{n+1}^{3}})(1+\frac{\Delta t}{t_{s,n}})}{1+\Delta t(1/ (2t_{s,n})+1/(2t_{s,n+1})+\Delta t/(2t_{s,n}t_{s,n+1}))}
\end{equation}

\section{B. Particle Test Problems}
We did various tests for above three integrators.

\begin{figure}[ht]
\centering
\includegraphics[width=0.8\textwidth]{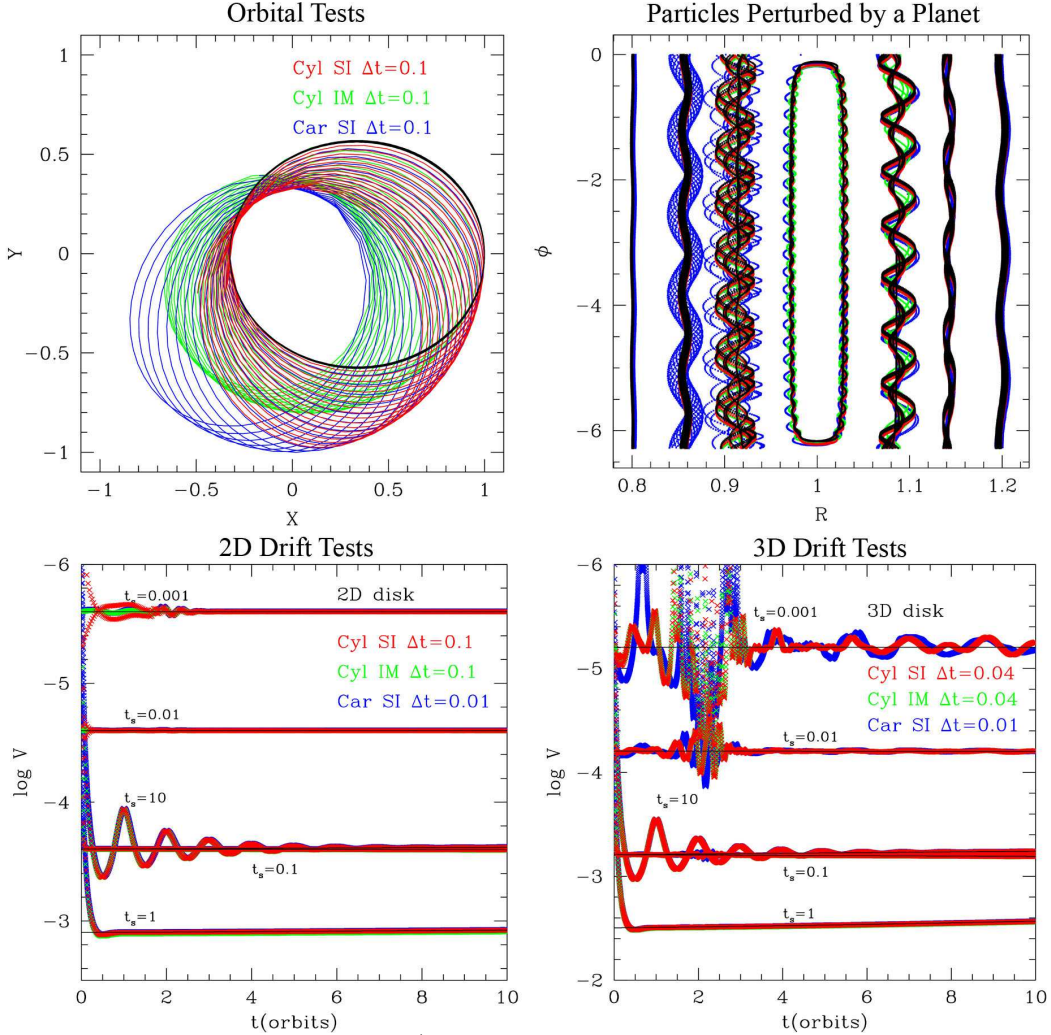} 
\vspace{-0.3 cm}
\caption{
Four test problems for particle integrators (Cartesian leapfrog: Car SI,
Cylindrical leapfrog: Cyl SI, and cylindrical fully implicit: Cyl IM. They are introduced in Appendix A, ).
The upper left panel: we drop one particle at $x=1$, $y=0$ with $v_{y}=0.7$ (corresponding
to $e=0.51$), and
follow the particle orbit using three integrators with $\Delta t=0.1$ for 20 orbits. The black ellipse is
the orbit calculated with Car SI and $\Delta t$=0.01. The upper right panel: with a planet
at $R=1$, $\phi=0$, we drop 8 particles uniformly from 0.8 to 1.2 and follow their orbits in the
corotating frame of the planet. The lower left panel: in a 2-D disk with $\Sigma=R^{-1}$, we 
drop particles with different stopping times and follow their inward drift. The black lines
are the analytical solutions based on Eq. \ref{eq:eqdrift}. The lower right panel: similar to
the lower left panel but the particles are in the midplane of a 3-D disk. } 
\label{fig:fig1}
\end{figure}

1. $Orbital$ $tests$ (upper left panel of Figure \ref{fig:fig1}): We release one particle at $x=1$, $y=0$ with $v_{y}=0.7$,
and integrate its orbit for 20 orbits.
The particle follows an eccentric orbit with $e=0.51$. The time steps are $\Delta t$=0.1, compared with
the orbital time ($2\pi$). All these orbits process since even symplectic integrators cannot simultaneously 
preserve angular momentum and energy exactly. However, both
Cyl SI (leapfrog in cylindrical coordinates) and Car SI (leapfrog in Cartesian coordinates)
preserve the geometric properties of the orbits (e.g. the shape of its orbit or its eccentricity), 
while Cyl IM leads to orbits with
decreasing eccentricity (the orbit becomes rounder with time).  
This is consistent with the phase-volume conserving properties of Cyl SI and Car SI.
Furthermore, Cyl SI outperforms Car SI with a slower procession rate.
The black orbit is calculated using Car SI with $\Delta t$=0.01 showing no visible orbital procession. 
Since $\Delta t$=0.01 is normally the time step used in our planet-disk simulations,
our integrators are quite accurate even if we integrate the orbit of particles having moderate eccentricity.  

2. $Perturbed$ $tests$ (upper right panel of Figure \ref{fig:fig1}):  We place a planet with the mass of $4\times 10^{-5}$ at $R=1$ and $\phi=0$, release
8 particles uniformly distributed from 0.8 to 1.2 , and follow their positions for 60 orbits in the frame corotating with the planet. 
Particle eccentricity is higher when they are closer to the planet. When they are very close to the
planet, they experience horseshoe orbits. Each color represents one integrator in the same way as Figure \ref{fig:fig1} (a). $\Delta t$ is 0.1. The black curve can be considered as the accurate orbit which is derived by using Car SI
with $\Delta t$=0.001. Again, Cyl SI follows the black curves very well.  The relative error of the Jacobi integral (J) for  the cases with $\Delta t$=0.1 is $|\Delta J|/J<10^{-4}$ even during close encounters.

\begin{figure*}[ht]
\centering
\includegraphics[width=0.8\textwidth]{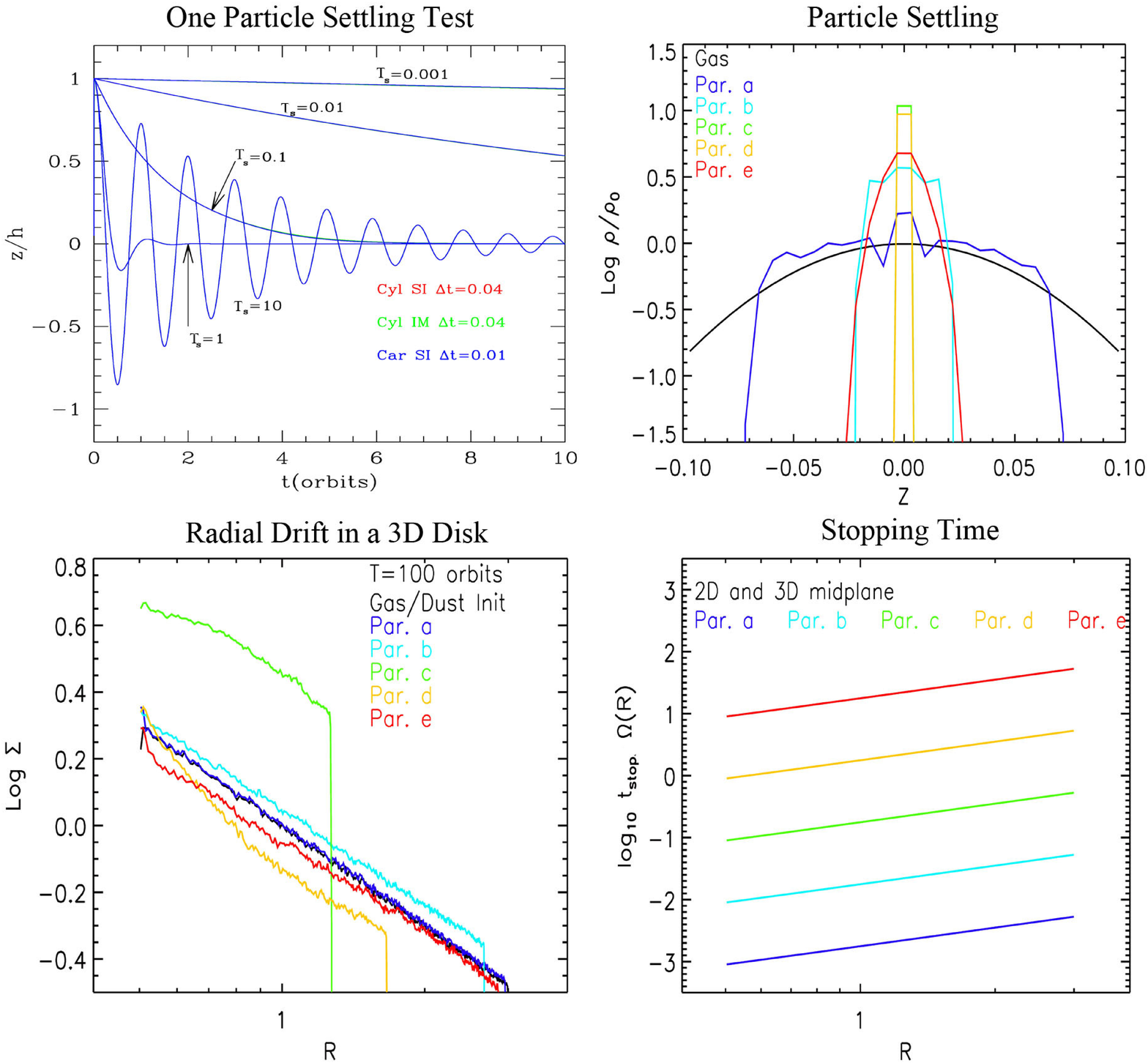}
\vspace{-1.2cm}
\caption{Various tests of particle settling and drift. The upper left panel:
five individual particles with different stopping times are dropped at one scale height above
the midplane. Their evolution is integrated with our three integrators. The particle with stopping time
unity settles fastest and particles with longer settling time oscillate around the midplane with decreasing amplitudes.
The upper right panel: particle density along $z$ at R=1 in the settling test at 10 orbits. Par. c and
 Par. d have settled to the midplane within one grid scale.The lower left panel: particle surface density 
 in the 3-D disk without planets at
 100 orbits.  Significant radial drift is observed for Par. c and Par. d. 
 The lower right panel: the dimensionless dust stopping time ($T_{s}=t_{s}\Omega(R)$) for Par. a to Par. e
 in 2-D and the midplane of 3-D disks in the initial condition.
}
\label{fig:fig3}
\end{figure*}

3. $2$-$D$ $Drift$ $test$ (lower left panel of Figure \ref{fig:fig1}): We set-up a 2-D gaseous disk with $\Sigma=R^{-1}$ and release 5 particles at $R=1$ with $t_{s}$ from 0.001
to 10. The radial domain is
$[0.5,3]$ with the resolution of 400.
Analytical calculation suggests that they will drift to the central star at speeds of
\begin{equation}
v_{R,d}=\frac{T_{s}^{-1}v_{R,g}-\eta v_{K,mid}}{T_{s}+T_{s}^{-1}}\,, \label{eq:eqvrd}
\end{equation}
where $v_{R,g}$ is the gas radial velocity, $v_{K,mid}$ is the midplane Keplerian velocity, and 
$\eta=-(R \Omega_{K}^{2}\rho_{g})^{-1}\partial P_{g}/\partial R$ is the ratio of the gas pressure gradient
to the stellar gravity in the radial direction (Weidenschilling 1977, Takeuchi \& Lin 2002). For this drift test, the gaseous disk is set to be in hydrostatic equilibrium with no radial velocity, so that Eq. \ref{eq:eqvrd} is
\begin{equation}
v_{R,d}=\frac{c_{s,mid}^{2}}{v_{K,mid}^{2}}\frac{ \partial {\rm ln} P}{\partial {\rm ln} R}\frac{v_{K,mid}}{T_{s}+T_{s}^{-1}}=\frac{ \partial {\rm ln} P}{\partial {\rm ln} R}\frac{(H/R)^{2}v_{K,mid}}{T_{s}+T_{s}^{-1}}\label{eq:eqdrift}
\end{equation}
For our isothermal 2-D disk set-up, $P\propto \Sigma\propto R^{-1}$ and $\partial {\rm ln}P /\partial {\rm ln} R=-1$.
$H/R$, $v_{K,mid}$, and $T_{s}$ are both functions of $R$.
The analytical solutions are drawn as the black solid lines.  Our integrators perform very well for all the particles.
However, to get the same accuracy, Car SI uses a time step which is one order of magnitude smaller due to
its inaccurate position at the Kick step (details in Appendix A). For the smallest stopping time, some numerical
instability was observed for Cyl SI at $t<2$ orbits, which is expected since the stopping time is already 2 orders of magnitude smaller than the numerical
time step. The fully implicit scheme (Cyl IM) performs very well in this case. Thus, for the smallest particles
in our simulations, we use the Cyl IM integrator.

4. $3$-$D$ $Drift$ $test$ (lower right panel of Figure \ref{fig:fig1}): We set-up a 3-D gaseous disk in the same way as our planet-disk interaction simulations in
the main text. The $R\times Z$ domain sizes are $[0.5,3]\times[-0.1,0.1]$, with the resolution of $400\times 32$. 
Again, 5 particles with $t_{s}$ from 0.001
to 10 are released at $R=1$ in the disk midplane. 
However, the disk is not in a perfect hydrostatic equilibrium initially and it has 
small amplitude velocity fluctuations at the magnitude of $v\sim 10^{-6}$. This affects the particle 
drift velocity and it suggests that numerical errors will dominate particle drift speeds for particles with $t_{s}<0.001$.
For the midplane in our 3-D disks, $P\propto \Sigma/H \propto R^{-2.5}$ and $\partial ln P/\partial ln R=-2.5$ in Eq. \ref{eq:eqdrift}. 
Again, the analytical drift velocities are the black lines. Our integrators give quite accurate drift speeds for these
particles.

We want to emphasize that particles drift faster in 3-D simulations than 2-D simulations by 
comparing Figure \ref{fig:fig1} (c) and (d). This is because in 2-D simulations, we cannot capture the vertical structure of the disk and always
use $\nabla (\Sigma c_{s}^{2})$ to represent the pressure gradient, while in 
 3-D disks the gas pressure gradient $\nabla (\rho c_{s}^{2})$ at the 
midplane is larger than $\nabla (\Sigma c_{s}^{2})$ so that it can lead to a faster drift speed.

5. $One$ $Particle$ $Settling$ $Test$ (upper left panel of Figure \ref{fig:fig3}): 
Particles with different stopping times are released at one scale height away from the midplane. 
If $T_{s}<1$, the particle vertical position decays exponentially towards the midplane on a timescale of $1/(T_{s}\Omega)$.
The fastest settling is achieved when $T_{s}\sim 1$.
If $T_{s}>1$, the particle oscillates periodically with the exponentially decaying amplitude on a timescale
of $2T_{s}/\Omega$ (Weidenschilling 1980, Nakagawa \etal 1986, Garaud \etal 2004). 

6. $Settling$ $Test$ $For$ $The$ $Particle$ $Disk$:
Since the settling timescale is
$t_{settle}\sim \Omega^{-1}(T_{s}+T_{s}^{-1})$ ($test$ 5), 
all particles in our sample (Par. a to f in Table 2) should settle to the midplane on a timescale of 1000 orbits. 
To demonstrate this point and study how particles collectively settle in disks,
we run a 3-D simulation with the initial dust distribution having the same scale height as  the gaseous disk scale height ($H_{p}\sim H$).
After 10 orbits, the azimuthally averaged particle density at $R=1$ along $z$ is shown in the upper right panel of Figure \ref{fig:fig3}.   
The $R-Z$ simulation domain has the size of $[0.5,3]\times[-0.25,0.25]$ with the resolution of $400\times 80$. 
There are $10^6$ particles for each particle type.
The smallest particles (Par. a) couple to the gas quite well so that the settling is relatively weak. 
For bigger particles (Pars. b and c) the settling is more significant. All type c particles have settled
to the midplane within one numerical grid at the midplane. Based on Table 2 and the lower right panel of Figure \ref{fig:fig3}, we can see that type c\&d particles have 
stopping time close to 1 and settle fastest. When particles start to become decouple from the gas ($T_{s}>1$),  they 
oscillate around the midplane and settle slowly (Par. e).

We learn two things from this test: 1) The particles we considered have settled to the midplane quite quickly. Thus we choose $H_{p}=0.2 H$ for all other simulations presented
in the main text.
2) Our numerical resolution is too low to correctly study
dust feedback to the gas near disk midplane. Even if all particles settle within one grid at the midplane, the particle density  only increases  by a factor
of 10 (par. c in the upper right panel of Figure \ref{fig:fig3}), which is 10\% of the gas density. This is still far from the regime where 
particle feedback to the gas, leading to streaming instability (Johansen \& Youdin 2007, Bai \& Stone 2010b), 
starts to play an important role. In order to reveal such effects, 
we need at least a factor of 10 higher 
resolution, which is beyond the reach of current computation power.

7. $Drift$ $Test$ $For$ $The$ $Particle$ $Disk$:
After studying particle settling, we also did a longer 3-D run without planets to understand radial drift of a number of particles. The 
set-up is similar to the above settling test except that now $H_{p}=0.2 H$ and we run the simulation for 200 orbits.  
The $R-Z$ simulation domain has the size of $[0.5,3]\times[-0.1,0.1]$ with the resolution of $400\times 32$. 
There are $10^6$ particles for each particle type.
The dust surface density for different particle species at 100 orbits is shown
in the lower left panel of Figure \ref{fig:fig3}. Particles with type c and d
drift fastest since their stopping times are close to 1 and the drift timescale for particles having $T_{s}=1$ is $\sim (R/H)^{2}/\Omega$ as in $test$ $3\&4$ above.
Just within 100 orbits, these two types of particles 
drift from R=3 to R=1.5.
However, Par. c and d have quite different surface density distributions due to the change of their dust stopping time 
($T_{s}$ are shown in the lower right panel of Figure \ref{fig:fig3}, and Eq. \ref{eq:TsR}). For Par. c, $T_{s}$ is close to 1 at the outer disk, thus particles at the outer disk
drift faster than the particles in the inner disk and the dust surface density piles up. For Par. d, the particles drift faster at the inner disk with $T_{s}\sim 1$ there
and the disk
loses mass quickly through the inner boundary and the surface density does not pile up\footnote{Strictly speaking, whether the dust surface density increases
or decreases depends on the sign of $d(R\Sigma_{d} v_{R})/dR$ in the particle continuity equation. 
In the Epstein regime and constant $c_{s}$, $v_{R}\propto \Sigma_{g}/(s v_{K})$ if $T_{s}\gg 1$ everywhere
and  $v_{R}\propto s/  (\Sigma_{g} v_{K}$) if $T_{s}\ll 1$ everywhere. Thus, assuming $\Sigma_{s}\sim \Sigma_{d}$ initially, the dust surface density increases for $T_{s}\ll 1$ 
and  decreases for $T_{s}\gg 1$.}. Through this test, we learn that the well-known fast radial drift problem limits our simulations to hundreds of orbits
if we want all types of particles to remain in the disk. 

\section{C. Hydro Planet-Disk Interaction Tests}
Since Athena 
has not been used to study planet-disk interaction using cylindrical grids before,
we compare its results with Fargo, which uses a finite difference scheme similar to ZEUS
 (Stone \& Norman 1992)
and the orbital advection scheme (Masset 2000).  Fargo is widely used
and tested, e.g. de Val-Borro et al. (2006).

\begin{figure}[ht]
\centering
\includegraphics[width=0.8\textwidth]{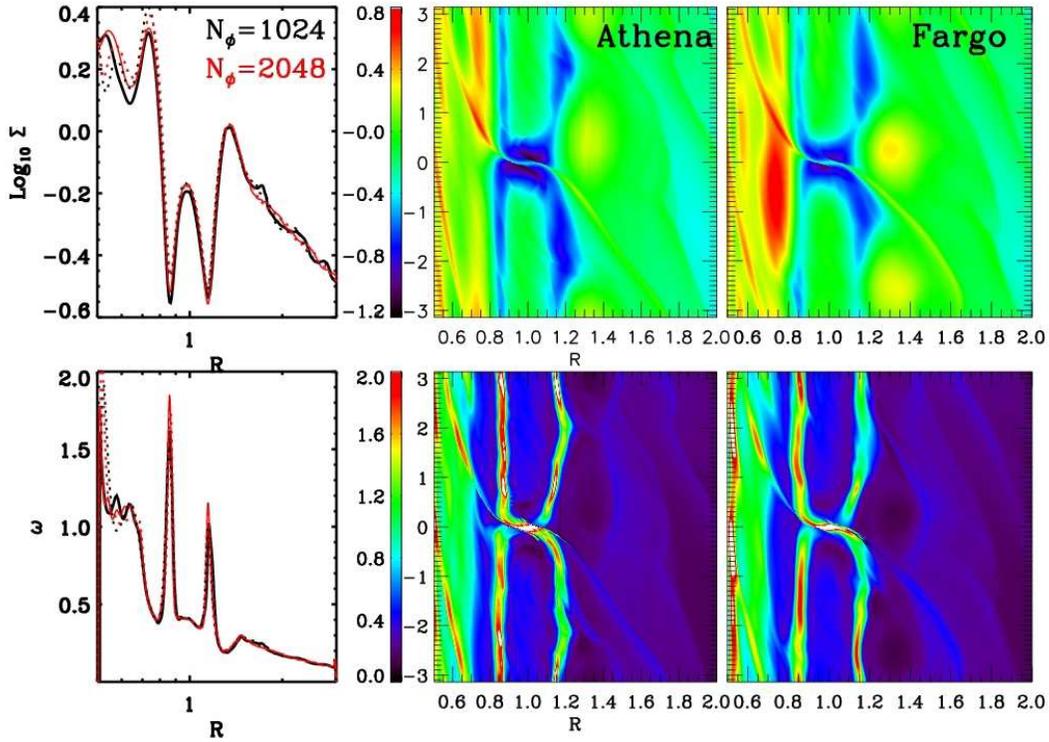} 
\vspace{-0.2 cm}
\caption{The disk surface density (upper panels)
and vorticity (lower panels) at 10 orbits in 2-D simulations with  a 5 $M_{th}$ planet using both Athena and Fargo. The left panels: The black curves are the azimuthally averaged surface density for low resolutions runs: M50D2A (using
Athena, the solid curves)
and  M50D2F (using Fargo, dotted curves). The red curves are from higher resolution runs (M50D2AH: solid curves and M50D2FH: dotted curves). The middle and right panels are the 2-D contours for both density and
vorticity for M50D2A and M50D2F. Good agreements have been achieved between different codes and different
resolutions.
} \label{fig:fig20}
\end{figure}

We set-up the Fargo simulation in the same way as our M50D2 case using the non-reflection boundary
condition. After 100 orbits, the disk surface density and vorticity are shown in Figure \ref{fig:fig20}.
Both codes give very similar results. We even double the resolution in both codes and
they give the same results (in the leftmost panels of Figure \ref{fig:fig20}). 

Athena is also capable of solving 3-D problems with magnetic fields and the radiative transfer, which
will be the focus of our upcoming papers.


\clearpage
\begin{thebibliography}

\bibitem[Andrews et al.(2011)]{2011ApJ...732...42A} Andrews, S.~M., Wilner, 
D.~J., Espaillat, C., et al.\ 2011, \apj, 732, 42 

\bibitem[Ataiee et 
al.(2013)]{2013A&A...553L...3A} Ataiee, S., Pinilla, P., Zsom, A., et al.\ 2013, \aap, 553, L3 



\bibitem[Ayliffe et al.(2012)]{2012MNRAS.423.1450A} Ayliffe, B.~A., Laibe, 
G., Price, D.~J., \& Bate, M.~R.\ 2012, \mnras, 423, 1450 


\bibitem[Bai 
\& Stone(2010a)]{2010ApJS..190..297B} Bai, X.-N., \& Stone, J.~M.\ 2010, \apjs, 190, 297


\bibitem[Bai 
\& Stone(2010b)]{2010ApJ...722.1437B} Bai, X.-N., \& Stone, J.~M.\ 2010, \apj, 722, 1437 



\bibitem[Barge 
\& Sommeria(1995)]{1995A&A...295L...1B} Barge, P., \& Sommeria, J.\ 1995, \aap, 295, L1 


\bibitem[Binney 
\& Tremaine(2008)]{2008gady.book.....B} Binney, J., \& Tremaine, S.\ 2008, Galactic Dynamics: Second Edition, by James Binney and Scott Tremaine.~ISBN 978-0-691-13026-2 (HB).~Published by Princeton University Press, Princeton, NJ USA, 2008.,  


\bibitem[Birnstiel et 
al.(2013)]{2013A&A...550L...8B} Birnstiel, T., Dullemond, C.~P., \& Pinilla, P.\ 2013, \aap, 550, L8 


\bibitem[Bitsch et 
al.(2013)]{2013A&A...549A.124B} Bitsch, B., Crida, A., Morbidelli, A., Kley, W., \& Dobbs-Dixon, I.\ 2013, \aap, 549, A124 

\bibitem[Colella 
\& Woodward(1984)]{1984JCoPh..54..174C} Colella, P., \& Woodward, P.~R.\ 1984, Journal of Computational Physics, 54, 174

\bibitem[Colella(1990)]{1990JCoPh..87..171C} Colella, P.\ 1990, Journal of 
Computational Physics, 87, 171  

\bibitem[Cuzzi 
\& Scargle(1985)]{1985ApJ...292..276C} Cuzzi, J.~N., \& Scargle, J.~D.\ 1985, \apj, 292, 276 


\bibitem[de Val-Borro et al.(2006)]{2006MNRAS.370..529D} de Val-Borro, M., 
Edgar, R.~G., Artymowicz, P., et al.\ 2006, \mnras, 370, 529 


\bibitem[de Val-Borro et 
al.(2007)]{2007A&A...471.1043D} de Val-Borro, M., Artymowicz, P., D'Angelo, G., \& Peplinski, A.\ 2007, \aap, 471, 1043 


\bibitem[Dong et al.(2011a)]{2011ApJ...741...56D} Dong, R., Rafikov, R.~R., 
Stone, J.~M., \& Petrovich, C.\ 2011, \apj, 741, 56 

\bibitem[Dong et al.(2011b)]{2011ApJ...741...57D} Dong, R., Rafikov, R.~R., 
\& Stone, J.~M.\ 2011, \apj, 741, 57 

\bibitem[Dong et al.(2012)]{2012ApJ...750..161D} Dong, R., Rafikov, R., 
Zhu, Z., et al.\ 2012, \apj, 750, 161 


\bibitem[Duffell 
\& MacFadyen(2012)]{2012ApJ...755....7D} Duffell, P.~C., \& MacFadyen, A.~I.\ 2012, \apj, 755, 7 

\bibitem[Espaillat et al.(2012)]{2012ApJ...747..103E} Espaillat, C., 
Ingleby, L., Hern{\'a}ndez, J., et al.\ 2012, \apj, 747, 103 


\bibitem[Forest(2006)]{2006JPhA...39.5321F} Forest, {\'E}.\ 2006, Journal 
of Physics A Mathematical General, 39, 5321 

\bibitem[Fouchet et 
al.(2007)]{2007A&A...474.1037F} Fouchet, L., Maddison, S.~T., Gonzalez, J.-F., \& Murray, J.~R.\ 2007, \aap, 474, 1037 


\bibitem[Fouchet et 
al.(2010)]{2010A&A...518A..16F} Fouchet, L., Gonzalez, J.-F., \& Maddison, S.~T.\ 2010, \aap, 518, A16 


\bibitem[Garaud et al.(2004)]{2004ApJ...603..292G} Garaud, P., 
Barri{\`e}re-Fouchet, L., \& Lin, D.~N.~C.\ 2004, \apj, 603, 292 

\bibitem[Gardiner 
\& Stone(2005)]{2005JCoPh.205..509G} Gardiner, T.~A., \& Stone, J.~M.\ 2005, Journal of Computational Physics, 205, 509 

\bibitem[Gardiner 
\& Stone(2008)]{2008JCoPh.227.4123G} Gardiner, T.~A., \& Stone, J.~M.\ 2008, Journal of Computational Physics, 227, 4123 

\bibitem[Goodman 
\& Rafikov(2001)]{2001ApJ...552..793G} Goodman, J., \& Rafikov, R.~R.\ 2001, \apj, 552, 793 

\bibitem[Goldreich 
\& Tremaine(1979)]{1979ApJ...233..857G} Goldreich, P., \& Tremaine, S.\ 1979, \apj, 233, 857 

\bibitem[Hedman et al.(2013)]{2013Icar..223..252H} Hedman, M.~M., Burns, 
J.~A., Hamilton, D.~P., \& Showalter, M.~R.\ 2013, ICARUS, 223, 252 

\bibitem[Heng 
\& Kenyon(2010)]{2010MNRAS.408.1476H} Heng, K., \& Kenyon, S.~J.\ 2010, \mnras, 408, 1476 

\bibitem[Hsieh 
\& Gu(2012)]{2012ApJ...760..119H} Hsieh, H.-F., \& Gu, P.-G.\ 2012, \apj, 760, 119 

\bibitem[Isella et al.(2013)]{2013ApJ...775...30I} Isella, A., P{\'e}rez, 
L.~M., Carpenter, J.~M., et al.\ 2013, \apj, 775, 30 


\bibitem[Johansen et 
al.(2004)]{2004A&A...417..361J} Johansen, A., Andersen, A.~C., \& Brandenburg, A.\ 2004, \aap, 417, 361 


\bibitem[Johansen et al.(2006)]{2006ApJ...643.1219J} Johansen, A., Henning, 
T., \& Klahr, H.\ 2006, \apj, 643, 1219 


\bibitem[Johansen 
\& Lacerda(2010)]{2010MNRAS.404..475J} Johansen, A., \& Lacerda, P.\ 2010, \mnras, 404, 475 


\bibitem[Johansen 
\& Youdin(2007)]{2007ApJ...662..627J} Johansen, A., \& Youdin, A.\ 2007, \apj, 662, 627 


\bibitem[Kley 
\& Nelson(2012)]{2012ARA&A..50..211K} Kley, W., \& Nelson, R.~P.\ 2012, \araa, 50, 211 

\bibitem[AKoller et al.(2003)]{2003ApJ...596L..91K} Koller, J., Li, H., 
\& Lin, D.~N.~C.\ 2003, \apjl, 596, L91 

\bibitem[Lambrechts 
\& Johansen(2012)]{2012A&A...544A..32L} Lambrechts, M., \& Johansen, A.\ 2012, \aap, 544, A32 

\bibitem[Lesur 
\& Papaloizou(2009)]{2009A&A...498....1L} Lesur, G., \& Papaloizou, J.~C.~B.\ 2009, \aap, 498, 1 


\bibitem[ALi et al.(2000)]{2000ApJ...533.1023L} Li, H., Finn, J.~M., 
Lovelace, R.~V.~E., \& Colgate, S.~A.\ 2000, \apj, 533, 1023 

\bibitem[Li et al.(2001)]{2001ApJ...551..874L} Li, H., Colgate, S.~A., 
Wendroff, B., \& Liska, R.\ 2001, \apj, 551, 874 


\bibitem[ALi et al.(2005)]{2005ApJ...624.1003L} Li, H., Li, S., Koller, J., 
et al.\ 2005, \apj, 624, 1003 

\bibitem[ALi et al.(2009)]{2009ApJ...690L..52L} Li, H., Lubow, S.~H., Li, 
S., \& Lin, D.~N.~C.\ 2009, \apjl, 690, L52 


\bibitem[ALin(2012a)]{2012ApJ...754...21L} Lin, M.-K.\ 2012, \apj, 754, 21 

\bibitem[ALin(2012b)]{2012MNRAS.426.3211L} Lin, M.-K.\ 2012, \mnras, 426, 
3211 

\bibitem[ALin(2013)]{2013ApJ...765...84L} Lin, M.-K.\ 2013, \apj, 765, 84 


\bibitem[ALin 
\& Papaloizou(1986)]{1986ApJ...309..846L} Lin, D.~N.~C., \& Papaloizou, J.\ 1986, \apj, 309, 846 

\bibitem[Lin 
\& Papaloizou(1993)]{1993prpl.conf..749L} Lin, D.~N.~C., \& Papaloizou, J.~C.~B.\ 1993, Protostars and Planets III, 749 

\bibitem[Lin 
\& Papaloizou(2010)]{2010MNRAS.405.1473L} Lin, M.-K., \& Papaloizou, J.~C.~B.\ 2010, \mnras, 405, 1473 


\bibitem[ALovelace et al.(1999)]{1999ApJ...513..805L} Lovelace, R.~V.~E., 
Li, H., Colgate, S.~A., \& Nelson, A.~F.\ 1999, \apj, 513, 805 

\bibitem[ALyra et 
al.(2009a)]{2009A&A...493.1125L} Lyra, W., Johansen, A., Klahr, H., \& Piskunov, N.\ 2009, \aap, 493, 1125 

\bibitem[ALyra et 
al.(2009b)]{2009A&A...497..869L} Lyra, W., Johansen, A., Zsom, A., Klahr, H., \& Piskunov, N.\ 2009, \aap, 497, 869 


\bibitem[Lyra 
\& Lin(2013)]{2013arXiv1307.3770L} Lyra, W., \& Lin, M.-K.\ 2013, arXiv:1307.3770 


\bibitem[ALyra 
\& Mac Low(2012)]{2012ApJ...756...62L} Lyra, W., \& Mac Low, M.-M.\ 2012, \apj, 756, 62 



\bibitem[AMasset(2000)]{2000A&AS..141..165M} Masset, F.\ 2000, \aaps, 141, 165

\bibitem[AMeheut et 
al.(2010)]{2010A&A...516A..31M} Meheut, H., Casse, F., Varniere, P., \& Tagger, M.\ 2010, \aap, 516, A31 

\bibitem[Meheut et al.(2012)]{2012MNRAS.422.2399M} Meheut, H., Yu, C., 
\& Lai, D.\ 2012, \mnras, 422, 2399 

\bibitem[AMeheut et 
al.(2012)]{2012A&A...545A.134M} Meheut, H., Meliani, Z., Varniere, P., \& Benz, W.\ 2012, \aap, 545, A134 

\bibitem[Morbidelli 
\& Nesvorny(2012)]{2012A&A...546A..18M} Morbidelli, A., \& Nesvorny, D.\ 2012, \aap, 546, A18 




\bibitem[M{\"u}ller et 
al.(2012)]{2012A&A...541A.123M} M{\"u}ller, T.~W.~A., Kley, W., \& Meru, F.\ 2012, \aap, 541, A123 

\bibitem[Murray(1994)]{1994Icar..112..465M} Murray, C.~D.\ 1994, ICARUS, 
112, 465 

\bibitem[Murray 
\& Dermott(1999)]{1999ssd..book.....M} Murray, C.~D., \& Dermott, S.~F.\ 1999, Solar system dynamics by Murray, C.~D., 1999,  


\bibitem[AMuto et al.(2010)]{2010ApJ...724..448M} Muto, T., Suzuki, T.~K., 
\& Inutsuka, S.-i.\ 2010, \apj, 724, 448 


\bibitem[ANakagawa et al.(1986)]{1986Icar...67..375N} Nakagawa, Y., Sekiya, 
M., \& Hayashi, C.\ 1986, ICARUS, 67, 375 

\bibitem[Ogilvie 
\& Lubow(2002)]{2002MNRAS.330..950O} Ogilvie, G.~I., \& Lubow, S.~H.\ 2002, \mnras, 330, 950 


\bibitem[Ormel 
\& Klahr(2010)]{2010A&A...520A..43O} Ormel, C.~W., \& Klahr, H.~H.\ 2010, \aap, 520, A43 


\bibitem[Paardekooper(2007)]{2007A&A...462..355P} Paardekooper, S.-J.\ 2007, \aap, 462, 355 


\bibitem[APaardekooper 
\& Mellema(2004)]{2004A&A...425L...9P} Paardekooper, S.-J., \& Mellema, G.\ 2004, \aap, 425, L9 


\bibitem[APaardekooper 
\& Mellema(2006)]{2006A&A...453.1129P} Paardekooper, S.-J., \& Mellema, G.\ 2006, \aap, 453, 1129 

\bibitem[Rafikov(2001)]{2001AJ....122.2713R} Rafikov, R.~R.\ 2001, \aj, 
122, 2713 


\bibitem[ARafikov(2002a)]{2002ApJ...572..566R} Rafikov, R.~R.\ 2002, \apj, 
572, 566 

\bibitem[Rafikov(2002b)]{2002ApJ...572..566R} Rafikov, R.~R.\ 2002, \apj, 
572, 566 




\bibitem[ARein 
\& Tremaine(2011)]{2011MNRAS.415.3168R} Rein, H., \& Tremaine, S.\ 2011, \mnras, 415, 3168 


\bibitem[ASkinner 
\& Ostriker(2010)]{2010ApJS..188..290S} Skinner, M.~A., \& Ostriker, E.~C.\ 2010, \apjs, 188, 290 

\bibitem[ASorathia et al.(2012)]{2012ApJ...749..189S} Sorathia, K.~A., 
Reynolds, C.~S., Stone, J.~M., \& Beckwith, K.\ 2012, \apj, 749, 189 

\bibitem[AStone et al.(2008)]{2008ApJS..178..137S} Stone, J.~M., Gardiner, 
T.~A., Teuben, P., Hawley, J.~F., \& Simon, J.~B.\ 2008, \apjs, 178, 137 

\bibitem[AStone 
\& Norman(1992)]{1992ApJS...80..753S} Stone, J.~M., \& Norman, M.~L.\ 1992, \apjs, 80, 753 

\bibitem[Tagger 
\& Pellat(1999)]{1999A&A...349.1003T} Tagger, M., \& Pellat, R.\ 1999, \aap, 349, 1003 

\bibitem[Tanaka 
\& Ida(1996)]{1996Icar..120..371T} Tanaka, H., \& Ida, S.\ 1996, ICARUS, 120, 371 

\bibitem[Tanaka 
\& Ida(1997)]{1997Icar..125..302T} Tanaka, H., \& Ida, S.\ 1997, ICARUS, 125, 302 

\bibitem[Takeuchi 
\& Lin(2002)]{2002ApJ...581.1344T} Takeuchi, T., \& Lin, D.~N.~C.\ 2002, \apj, 581, 1344 


\bibitem[ATanaka et al.(2002)]{2002ApJ...565.1257T} Tanaka, H., Takeuchi, 
T., \& Ward, W.~R.\ 2002, \apj, 565, 1257 

\bibitem[van der Marel et al.(2013)]{2013Sci...340.1199V} van der Marel, 
N., van Dishoeck, E.~F., Bruderer, S., et al.\ 2013, Science, 340, 1199 


\bibitem[AVarni{\`e}re 
\& Tagger(2006)]{2006A&A...446L..13V} Varni{\`e}re, P., \& Tagger, M.\ 2006, \aap, 446, L13 

\bibitem[Ward 
\& Hourigan(1989)]{1989ApJ...347..490W} Ward, W.~R., \& Hourigan, K.\ 1989, \apj, 347, 490 


\bibitem[AWard(2009)]{2009LPI....40.1477W} Ward, W.~R.\ 2009, Lunar and 
Planetary Institute Science Conference Abstracts, 40, 1477 

\bibitem[AWeidenschilling(1977)]{1977MNRAS.180...57W} Weidenschilling, 
S.~J.\ 1977, \mnras, 180, 57 

\bibitem[AWeidenschilling(1980)]{1980Icar...44..172W} Weidenschilling, 
S.~J.\ 1980, ICARUS, 44, 172


\bibitem[AWhipple(1972)]{1972fpp..conf..211W} Whipple, F.~L.\ 1972, From 
Plasma to Planet, 211 

\bibitem[AWisdom 
\& Holman(1992)]{1992AJ....104.2022W} Wisdom, J., \& Holman, M.\ 1992, \aj, 104, 2022 

\bibitem[AYoshida(1993)]{1993CeMDA..56...27Y} Yoshida, H.\ 1993, Celestial 
Mechanics and Dynamical Astronomy, 56, 27 


\bibitem[Youdin(2010)]{2010EAS....41..187Y} Youdin, A.~N.\ 2010, EAS 
Publications Series, 41, 187

\bibitem[AYoudin 
\& Lithwick(2007)]{2007Icar..192..588Y} Youdin, A.~N., \& Lithwick, Y.\ 2007, ICARUS, 192, 588 



\bibitem[Yu et al.(2010)]{2010ApJ...712..198Y} Yu, C., Li, H., Li, S., 
Lubow, S.~H., \& Lin, D.~N.~C.\ 2010, \apj, 712, 198 


\bibitem[Yu 
\& Lai(2013)]{2013MNRAS.429.2748Y} Yu, C., \& Lai, D.\ 2013, \mnras, 429, 2748 


\bibitem[AZhu et al.(2012)]{2012ApJ...755....6Z} Zhu, Z., Nelson, R.~P., 
Dong, R., Espaillat, C., \& Hartmann, L.\ 2012, \apj, 755, 6 


\bibitem[AZhu et al.(2013)]{2013ApJ...768..143Z} Zhu, Z., Stone, J.~M., 
\& Rafikov, R.~R.\ 2013, \apj, 768, 143 



\end{thebibliography}
\end{document}